\documentclass[acmsmall]{acmart}

\usepackage{tikz}
\usepackage{amsmath}
\usepackage{filecontents}
\usepackage[flushleft]{threeparttable}
\usepackage{float}
\usepackage{amsmath,amsfonts}

\usepackage{graphicx}
\usepackage{textcomp}

 \usepackage{url}

\usepackage{lipsum,tabularx}
\usepackage{wrapfig}
\usepackage{multicol}
\usepackage{multirow}

\usepackage{colortbl}
\usepackage{booktabs}
\usepackage{setspace}

\def\Snospace~{\S{}}

\usepackage[T1]{fontenc}
\usepackage{pifont}

\usepackage{balance}

\usepackage{bm}
\usepackage{fp}
\usepackage{siunitx}
\usepackage{amsthm}

\usepackage[labelfont=bf,font=small,skip=5pt]{caption}

\usepackage{comment}

\usepackage{fancyhdr}
\usepackage{tikz}

\usepackage{xspace}

\newcommand{\boxbeg}{
  \vspace{2px}
  \noindent\begin{tabular}{|l|}\hline
    \begin{minipage}{3.2in}
      \vspace{2px}
      \noindent
      }

      \newcommand{\boxend}{
      \vspace{2px}
    \end{minipage} \\ \hline
  \end{tabular}
  \vspace{-10pt}
}
\usepackage{listings}

\usepackage[normalem]{ulem}
\useunder{\uline}{\ul}{}

\newcommand{\updmn}[2]{#2}
\newcommand{\upd}[2]{#2}
\newcommand{\updd}[2]{#2}

\newcommand{\eg}[0]{\textit{e.g.}}
\newcommand{\ie}[0]{\textit{i.e.}}
\newcommand{\et}[0]{\textit{et al.}}

\AtBeginDocument{%
  }

\setcopyright{acmlicensed}
\acmJournal{CSUR}
\acmYear{2025} \acmVolume{1} \acmNumber{1} \acmArticle{1} \acmMonth{1}
\acmDOI{10.1145/3716497}
\begin{document}

\title{Deep Learning Library Testing: Definition, Methods and Challenges}

\author{Xiaoyu Zhang}
\email{zxy0927@stu.xjtu.edu.cn}
\orcid{0000-0001-7010-6749}
\affiliation{
  \institution{School of Cyber Science and Engineering, Xi'an Jiaotong University}
  \city{Xi'an}
  \country{China}
}

\author{Weipeng Jiang}
\email{lenijwp@stu.xjtu.edu.cn}
\orcid{0000-0002-0382-6401}
\affiliation{
  \institution{School of Cyber Science and Engineering, Xi'an Jiaotong University}
  \city{Xi'an}
  \country{China}
}

\author{Chao Shen}
\email{chaoshen@mail.xjtu.edu.cn}
\orcid{0000-0002-6959-0569}
\authornote{Corresponding author}
\affiliation{
  \institution{School of Cyber Science and Engineering, Xi'an Jiaotong University}
  \city{Xi'an}
  \country{China}
}

\author{Qi Li}
\email{qli01@tsinghua.edu.cn}
\orcid{0000-0001-8776-8730}
\affiliation{
  \institution{Institute for Network Sciences and Cyberspace, Tsinghua University}
  \state{Beijing}
  \country{China}
}

\author{Qian Wang}
\email{qianwang@whu.edu.cn}
\orcid{0000-0002-8967-8525}
\affiliation{
  \institution{School of Cyber Science and Engineering, Wuhan University}
  \city{Wuhan}
  \country{China}
}

\author{Chenhao Lin}
\email{linchenhao@xjtu.edu.cn}
\orcid{0000-0002-6265-7345}
\affiliation{
  \institution{School of Cyber Science and Engineering, Xi'an Jiaotong University}
  \city{Xi'an}
  \country{China}
}

\author{Xiaohong Guan}
\email{xhguan@mail.xjtu.edu.cn}
\orcid{0000-0002-8826-0362}
\affiliation{
  \institution{School of Cyber Science and Engineering, Xi'an Jiaotong
  University}
  \city{Xi'an}
  \country{China}
}

\renewcommand{\shortauthors}{Zhang and Jiang, et al.}

\begin{abstract}
Recently, software systems powered by deep learning (DL) techniques have significantly facilitated people's lives in many aspects.
\upd{Response to R3Q1: }{
As the backbone of these DL systems, various DL libraries undertake the underlying optimization and computation.
However, like traditional software, DL libraries are not immune to bugs.
These bugs may be propagated to programs and software developed based on DL libraries, thereby posing serious threats to users' personal property and safety.
}
Studying the characteristics of DL libraries, their associated bugs, and the corresponding testing methods is crucial for enhancing the security of DL systems and advancing the widespread application of DL technology.
This paper provides an overview of the testing research on various DL libraries, discusses the strengths and weaknesses of existing methods, and provides guidance and reference for the application of DL library testing methods.
This paper first introduces the workflow of DL underlying libraries and the characteristics of three kinds of DL libraries involved, namely DL framework, DL compiler, and DL hardware library.
Subsequently, this paper constructs a literature collection pipeline and comprehensively summarizes existing testing methods on these DL libraries to analyze their effectiveness and limitations.
It also reports findings and the challenges of existing DL library testing in real-world applications for future research.
\end{abstract}




\maketitle


\section{Introduction}



With the development of deep learning (DL) techniques, the DL systems that are driven by DL models have been applied in many fields, providing societal benefits in areas like image recognition~\cite{he2016deep}, self-driving~\cite{grigorescu2020survey}, and natural language processing~\cite{li2018deep}.
As the backbone of DL systems, the security and safety of the underlying DL library have received more and more attention.
The DL library (\eg, PyTorch) is responsible for performing specific computations for training or inference DL models and implementing optimized operations on DL hardware.
\upd{Response to R3Q1: }{
Developers are highly relying on DL libraries to develop software and systems containing DL components.}
Tesla relies on PyTorch which is one of the most popular DL libraries to solve problems related to the self-driving domain~\cite{tesla1pytorch}.
TensorFlow, which is another popular DL library, undertakes many important business tasks of Google, Intel, and other companies~\cite{google1tensorflow}.

\upd{Response to R3Q1: }{Similar to traditional software, the DL library also has bugs, which can be propagated to the DL systems developed upon these libraries and cause the systems to make erroneous predictions, generate huge overhead, and even crash~\cite{pham2019cradle,wei2022free}, thereby jeopardizing user property and personal safety.}
For example, in recent years, the self-driving systems developed by Tesla and Uber have experienced abnormal behaviors during driving and eventually led to fatal crashes~\cite{teslacase,ubercase}, which further arouse people's concerns about the security and safety of DL systems and the underlying libraries.


At present, researchers have proposed a series of tools and methods~\cite{pham2019cradle,wei2022free,deng2022fuzzing} to discover bugs such as crashes, overflows, and numerical errors on the DL libraries represented by DL frameworks (\eg, TensorFlow, PyTorch), aiming to guarantee the security and usability of the DL system built upon these libraries.
How to deeply understand the bugs of the underlying libraries of the DL system and design testing methods for these libraries needs to be solved urgently and is of great significance.
Although researchers have proposed various DL library testing methods, there are still many challenges.
Firstly, there are various types of DL libraries, including DL frameworks, DL compilers, etc.
Different DL libraries undertake different calculation and optimization functions, and there are significant differences between their inputs, outputs, and implementations.
As a result, existing testing methods are diverse and highly targeted to specific libraries, leading to a lack of general and systematic testing methods for different DL libraries and evaluation benchmarks for different testing methods.
Furthermore, existing research has a limited understanding of DL library bugs and mainly focuses on crashes and numerical errors.
They rarely evaluate and identify other bugs in DL libraries (\eg, performance bugs), which limits the effectiveness of these methods.
Therefore, it is significant to conduct induction, analysis, and discussion on the existing research in the field of DL library testing to find limitations and provide guidance for subsequent research directions in related fields.

\begin{table*}[]
\caption{\upd{Response to R2Q1: }{Comparison Between Our Survey and Related Papers.}}
\label{tab:intro_compare}
\centering
\scriptsize
\tabcolsep=2pt
\begin{tabular}{cccccccccc}
\toprule
\multirow{3}{*}{\textbf{Paper}} & \multirow{3}{*}{\textbf{Year}} & \multicolumn{5}{c}{\textbf{Covered Component}} & \multicolumn{2}{c}{\textbf{Covered Property}} & \multirow{3}{*}{\textbf{\begin{tabular}[c]{@{}c@{}}Discussion Across\\ Components\end{tabular}}} \\ \cmidrule(r){3-7} \cmidrule(r){8-9}
 &  & \textbf{Framework} & \textbf{Compiler} & \textbf{\begin{tabular}[c]{@{}c@{}}Hardware\\ Library\end{tabular}} & {\textit{Models}} & {\textit{\begin{tabular}[c]{@{}c@{}}DL-based \\ Software\end{tabular}}} & \textbf{Correctness} & \textbf{Efficiency} &  \\ \midrule
Braik \et~\cite{braiek2020testing} & 2018 & \checkmark &  &  & \checkmark & \checkmark & \checkmark & \checkmark &  \\
 Zhang \et~\cite{zhang2020machine} & 2020 & \checkmark &  &  & \checkmark & \checkmark & \checkmark & \checkmark &  \\
Li \et~\cite{li2020deep}& 2020 &  & \checkmark &  &  &  & \checkmark &  &  \\
Zhang \et~\cite{zhang2022testing} & 2022 &  &  &  & \checkmark & \checkmark & \checkmark & \checkmark &  \\
 Ma \et~\cite{ma2023survey} & 2023 & \checkmark &  &  &  & & \checkmark &  &  \\
 Hu \et~\cite{hu2023research} & 2023 & &  &  & \checkmark  & & \checkmark &  &  \\
 Ji \et~\cite{ji2023survey} & 2023 & \checkmark &  &  & & & \checkmark &  &  \\
Our Survey & 2024 & \checkmark & \checkmark & \checkmark &  &  & \checkmark & \checkmark & \checkmark \\ \bottomrule
\end{tabular}
\end{table*}

\upd{Response to R2Q1, R3Q2 and R3Q7: }{
However, existing surveys on the DL library testing are limited.
We have compared related surveys in~\autoref{tab:intro_compare} and observed that existing works mainly focus on testing and repairing DL models or software built upon the models~\cite{braiek2020testing,gezici2022systematic,martinez2022software,zhang2022testing,hu2023research} (italic columns in~\autoref{tab:intro_compare}).
Even though some works involve testing DL libraries, they typically only cover certain components of various libraries and fail to provide a detailed introduction, analysis, and discussion across testing methods for different DL library components.
For example, Zhang \et~\cite{zhang2020machine} and Ji \et~\cite{ji2023survey} discussed testing methods for DL frameworks, while Li \et~\cite {li2020deep} focused on the workflow and performance of DL compilers.
\updmn{Response to R3Q1: }{
To fill this gap, this paper focuses on the bugs that affect the functionality correctness and efficiency of DL libraries and comprehensively summarizes corresponding testing methods for three DL library components, namely \ding{182} \textbf{DL frameworks} that execute the DL program and construct DL model, \ding{183} \textbf{DL compilers} that compile and translate DL models into optimized operators, and \ding{184} \textbf{DL hardware libraries} that map operators to the hardware to perform calculations.}
This paper further analyzes the advantages and limitations of existing methods and delves into the challenges and future research opportunities in DL library testing.
Our goal is to provide a set of practical findings to promote the development of DL library testing research, thereby ensuring the security and reliability of DL systems built on these libraries in real-world scenarios.
Note that the scope of this paper includes the testing methods on three DL library components, but does not include the DL model and DL program testing.}
The main contribution can be summarized as follows:
\begin{itemize}
    
\item \updmn{Response to R2Q1: }{We present the first comprehensive and detailed survey of testing methods for various libraries in the DL workflow, including DL frameworks, compilers, and underlying hardware libraries.
Our work expands and enhances existing DL testing surveys, which only focus on the specific DL component (\eg, DL framework) and lack an in-depth analysis and discussion of bugs and testing methods for different DL library components.}
\item We propose a novel taxonomy in three test components to provide an accessible overview of works that focus on libraries at different stages in the DL workflow, namely DL framework, DL compiler, and DL hardware library, as shown in~\autoref{fig:overview}.
For each stage, we comprehensively summarize and present existing work according to their testing techniques and provide an in-depth analysis at the end to characterize some critical problems.
\item We provide a set of practical findings based on the literature survey and outline the main challenges and future research directions of DL library testing, aiming to promote the development of DL software security and safety.

\end{itemize}

This paper is organized as follows. \autoref{sec:pre} describes the workflow and three key components of the DL underlying libraries and other preliminary knowledge.
\autoref{sec:method} explains the methodology of our paper, the review questions to be answered, and the literature collection process.
Based on the review questions,~\autoref{sec:rq1},~\autoref{sec:framework},~\autoref{sec:dependency} and~\autoref{sec:hardware} introduce various testing methods for different DL library components, and discuss the advantages and limitations of these methods.
Then~\autoref{sec:future} provides practical findings from our survey and discusses existing challenges in DL library testing, and~\autoref{sec:conclusion} concludes this paper.

\section{Preliminary}
\label{sec:pre}

\subsection{DL Model}

\upd{Response tot R3Q4: }{
A machine learning (ML) model is a parameterized function \(F: X \mapsto Y\), where \(x\in X\) is an \(m-\)dimensional input and \(y\in Y\) is the corresponding output label.
As a family of the ML model, the DL model is a neural network typically composed of several connected layers.
An $n$-layered DL model can be represented as \(F_\theta =  l_1 \circ l_2 \circ \cdot\cdot\cdot \circ l_n\), where \(l\) represents a layer and \(\theta\) indicates the learnable parameters in the DL model.
The developers first need to train the DL model on the given data and update the model parameters \(\theta\) in the training progress.
Then, in the inference process, the trained DL model can predict the output for the given input (\eg, an image, or a sentence).}

The training process of a DL model consists of the \textit{forward propagation} stage and the \textit{backward propagation} stage, and the model inference process only uses the former.
The forward propagation stage calculates model output \(F_\theta(x_i)\) based on the input tensor \(x_i\) and initialized model parameter \(\theta\).
The backward propagation stage evaluates the difference between \(F_\theta(x_i)\) and the ground truth label \(y_i\) by a loss function \(\mathcal{L} (F_\theta(x),y)\) and updates model parameters \(\theta\) to minimize the value of \(\mathcal{L}\).
The forward propagation and backward propagation stages will be repeated until the training reaches the predetermined stopping criteria.
In the DL program, developers call the APIs provided by the DL framework to build and train a DL model and each layer \(l_i\) in the model can be directly constructed by one or several DL framework APIs.


\subsection{DL Library Components}
\label{sec:librarydefinition}

\begin{figure}
    \centering     
    \includegraphics[width=0.9\linewidth]{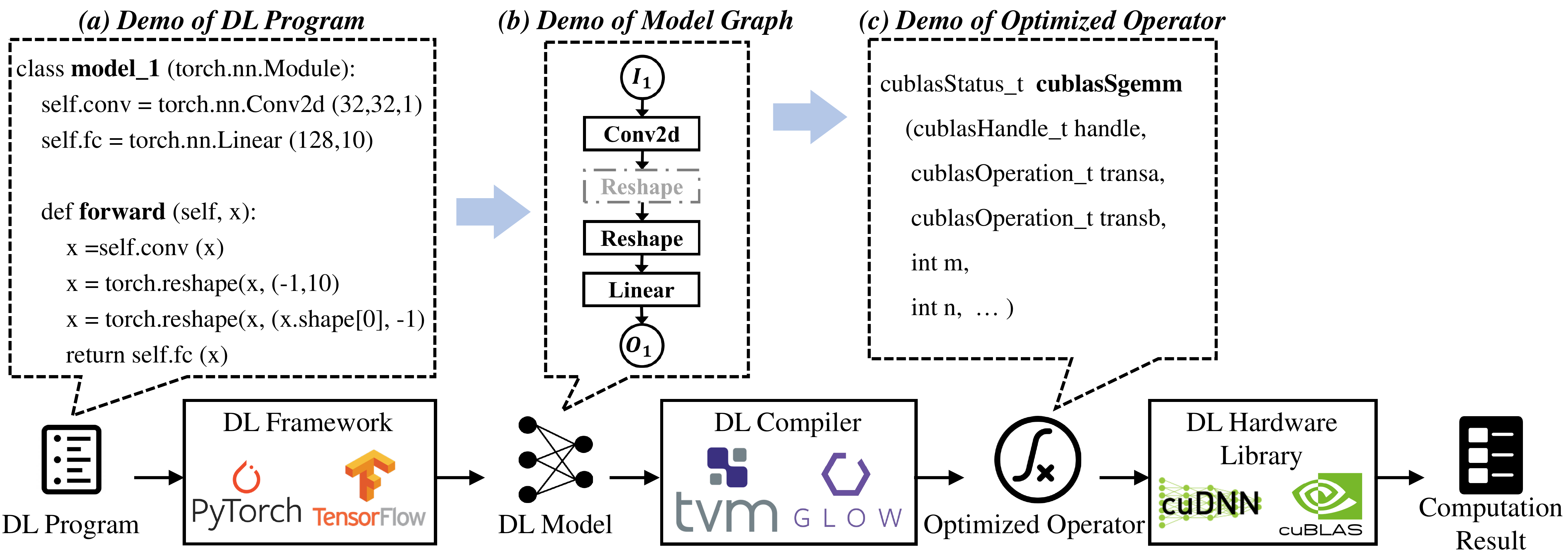}
    \caption{\updmn{Response to R2Q3: }{Overarching Workflow of the DL system and DL program}}\label{fig:overview}
\end{figure}

DL libraries implement the abstract DL model based on the program and perform specific operations and optimizations on the underlying hardware to obtain computation results of DL model training and inference.
Executing the DL program and building a DL model on the DL underlying libraries mainly involves three components, namely the DL framework, DL compiler, and DL hardware library.
\updmn{Response to R2Q3: }{
The three components are closely related, cooperate with each other, and together form the DL workflow, as shown in~\autoref{fig:overview}.
The code in dotted boxes presents the demo inputs of each component.
Developers first need to call the DL framework APIs to construct a DL program.
As shown in~\autoref{fig:overview} (a), the DL program calls the DL library APIs to define the convolutional layer (\texttt{torch.nn.Conv2d}), linear layer (\texttt{torch.nn.Linear}), and reshape operation (\texttt{torch.reshape}).
DL framework then executes the DL program and constructs the corresponding DL model.
\autoref{fig:overview} (b) shows the corresponding DL model graph.
Subsequently, the DL compiler translates the input model into intermediate representations (IRs) and performs optimization, and outputs the optimized operators and code.
During the optimization process, the DL compiler eliminates the `Reshape' in the gray dashed box of~\autoref{fig:overview} (b) due to functional redundancy.
It also reformulates `Conv2d' as the matrix multiplication operator `cublasSgemm' in~\autoref{fig:overview} (c).
Finally, the DL hardware library (\eg, cuDNN) accepts the output of the compiler and maps calculations to the DL hardware to perform calculations and obtain the results.
}

\subsubsection{DL Framework}

DL frameworks are high-level DL libraries that provide a user-friendly interface for users to conveniently design, train, and deploy DL models.
\upd{Response to R3Q5: }{
Industry and academia have proposed various DL frameworks, including TensorFlow~\cite{abadi2016tensorflow}, PyTorch~\cite{paszke2019pytorch}, etc.
TensorFlow supports a variety of program languages (\eg, C++, Python, and Go) and is currently one of the most popular DL frameworks.
PyTorch is rewritten and optimized based on the DL framework Torch. Nowadays, PyTorch and TensorFlow have active developer communities and are the most commonly used test objects in DL framework testing research.
Note that the libraries that extend the functionality of DL frameworks or empower DL development are not within the scope of this paper, such as ONNX which provides an open-sourced format for DL models~\cite{ONNX} and AutoML pipelines that reduce manual efforts in designing models~\cite{jin2019auto}.
}

Like traditional software, DL frameworks provide numerous flexible APIs to call functions, perform operations, and construct neural networks.
Taking PyTorch~\cite{paszke2019pytorch} as an example, its API includes performing basic matrix operations (\eg, \texttt{torch.mul} for multiply operation), calculating loss functions (\eg, \texttt{torch.nn.MSELoss} for measuring mean squared error), and building models layers (\eg, \texttt{torch.nn.Conv2d} for convolution layers).
After the developer calls these APIs in the program, the DL framework will build the corresponding abstract DL model.

\subsubsection{DL Compiler}

To reduce the burden of manually optimizing DL models on various DL hardware (\eg, TPU) and hardware libraries (\eg, cuDNN), researchers have developed the DL compiler~\cite{li2020deep}.
It takes the abstract model described by the DL framework as input, and then automatically optimizes and generates operators and codes as output to ensure that the DL hardware library can efficiently execute calculations of the DL model.
Therefore, DL compilers are generally closely related to and work together with the DL framework that builds abstract DL models.
The currently popular DL compilers include Glow~\cite{rotem2018glow}, TVM~\cite{chen2018tvm}, etc.
Glow is designed to implement state-of-the-art optimizations and generate code for neural network graphs.
TVM provides graph-level and operator-level optimizations for DL models, whose optimized performance on some hardware is competitive with state-of-the-art hand-tuned libraries.

Similar to traditional compilers, DL compilers implement a layered design, which mainly consists of the compiler frontend and the compiler backend.
The intermediate representation (IR), as the abstract of the program, exists in both the frontend and the backend~\cite{li2020deep}.
The front end converts the DL model from the DL framework into a computation graph and optimizes the graph with various methods (\eg, reducing redundancy).
In this process, IR is mainly used to express DL models, construct the control flow and dependencies between operators and data, etc.
For the computation graph, the backend performs hardware-specific optimizations by leveraging third-party tools and customizing compilation passes based on prior knowledge.
Finally, DL compilers convert the DL model into operators and code which can be used to perform calculations on the DL hardware and hardware libraries, optimizing inference execution speed and memory usage.

\subsubsection{DL Hardware Library}

Researchers have designed a variety of DL hardwares, such as CPU, GPU, and TPU, to apply DL techniques in different scenarios.
To adapt to given DL hardware and map the computation to DL hardware efficiently, researchers have developed a series of DL hardware libraries (\eg, cuDNN, cuBLAS) that implement optimized linear algebras, matrix multiplication, DL operators, etc.
For example, cuDNN~\cite{chetlur2014cudnn} is a DL accelerate library developed by NVIDIA, which is used to achieve high-performance computing on its developed GPU (\eg, RTX3090).
In addition, cuBLAS provides a basic linear algebra library for the computation on GPU.

DL hardware libraries select appropriate algorithms for different hardware and deployment environments, thus they can realize specific calculations on DL hardware.
They also call different operators according to the data type to speed up the operation of the model on the given data.
For example, cuBLAS implements various matrix-matrix multiplication operators (\eg, cublasHgemm, cublasSgemm) to map the calculations of different data types to the hardware.


\subsubsection{Uniqueness, Relationship, and Similarities}

\upd{Response to R1Q1: }{
\textbf{Uniqueness}: The three DL components separately have their unique characteristics.
Firstly, different components are in different abstraction levels of DL workflow and the levels decrease from frameworks to hardware libraries.
Moreover, to complete different tasks, they also have unique structural designs.
DL frameworks typically implement modular interfaces for different stages of model execution (\eg, forward propagation, and backward propagation).
DL compilers focus on optimizing and generating operators.
Similar to traditional compilers, DL compilers use a layered design, including frontend and backend to generate and optimize IRs.
The design of DL hardware libraries is related to their usage scenarios, hardware, and tasks, and there is no general structural design.
\textbf{Relationship}: Although DL frameworks, DL compilers, and DL hardware libraries have unique designs and abstraction levels, they have close relationships with each other and collaborate to form a DL workflow.
Among them, DL frameworks are user-centric and provide high-level abstractions for deep learning tasks.
Developers use DL framework APIs to design and construct DL models.
The DL framework relies on DL hardware libraries at the lowest level to execute specific computations of DL models.
DL hardware libraries provide highly optimized routines for specific hardware architectures.
DL compilers at the intermediate level bridge the gap between high-level frameworks and low-level hardware libraries.
It optimizes the DL models generated by DL frameworks to utilize DL hardware libraries for more efficient computation.
\textbf{Similarities}: In terms of similarity, all three DL library components share the same goal of enabling efficient and convenient DL operations.
In addition, with the rapid development of DL techniques, all three DL library components are evolving rapidly, which has laid hidden dangers and bugs in the functional correctness and efficiency of their implementation.
}


\subsection{DL Library Testing}

DL library testing aims to discover flaws and errors in the DL libraries.
The DL library bug is essentially a kind of software bug.
Referring to the prior research~\cite{zhang2020machine,IEEEstdBug}, 
we define the behavior that the actual function of the DL library does not meet the requirements and specifications as a DL library bug.

\begin{definition}[DL Library Bug]
    A DL library bug refers to any imperfection or deficiency in a DL library that causes the actual function performed by the DL library to fail to meet the expected requirements or specifications.
\end{definition}

Based on the above definition of DL library bugs, we define DL library testing as follows.

\begin{definition}[DL Library Testing]
    DL library testing refers to any activity designed to discover and identify DL library bugs.
\end{definition}

\upd{Response to R3Q2: }{
Note that this paper focuses on the test methods that detect functional and performance bugs on DL libraries.
In the literature collection, we have observed that existing research mainly tests and discovers bugs, with few works exploring security vulnerabilities and weaknesses.
We provide a discussion in~\autoref{sec:future}.
In addition, the scope of this paper does not include DL program testing and DL model testing.
The former aims to discover and repair errors in DL programs, rather than bugs in the DL library called by the programs~\cite{zhang2021autotrainer, wardat2021deeplocalize}, while the latter focuses on the security properties (\eg, robustness) and problems of the DL model itself~\cite{zhang2022testing}.
}

\section{Research Method}
\label{sec:method}
\upd{Response to R1Q1, R1Q4, R2Q2, R3Q3, and R3Q6: }{
In this paper, we followed the methodology of the Systematic Literature Reviews (SLR) to conduct research, which is proposed by Kitchenha~\cite{kitchenham2009systematic,kitchenham2022segress} and widely adopted in software engineering.
Guided by the methodology, we engaged in the SLR through the following three phases: 
\begin{enumerate}
    \item Planning: In this initial phase, the necessity for conducting a systematic literature review on Deep Learning Library Testing is established, and the goals and review questions of the review are clearly defined.
    \item Conducting: This phase involves the construction of a pipeline for searching relevant literature, followed by a step-by-step process of selection and quality assessment, culminating in the formation of a final pool of papers.
    \item Reporting: A detailed and comprehensive report tailored to the intended audience.
\end{enumerate}}

\upd{Response to R2Q2: }{
In this section, we will first introduce the specific processes of Planning ( in~\autoref{sec:sub_plan} ) and Conducting ( in~\autoref{sec:sub_conduct} ).
The detailed reporting results will be elaborated upon in subsequent sections.
}

\begin{table*}[]
\caption{Review Questions (RQs) in Our Survey}
\label{tab:rqs}
\centering
\scriptsize
\tabcolsep=5pt
\begin{tabular}{cl}
\toprule
\multicolumn{2}{c}{\textbf{Goal 1:  Characteristics of DL Library Testing}} \\ \midrule
RQ 1.1 & Which test properties and bugs are focused on in DL library testing?\\
RQ 1.2 & What are the commonly used testing techniques for DL libraries? \\
RQ 1.3 & What are the characteristics and differences between the testing on different DL library components? \\ \midrule
\multicolumn{2}{c}{\textbf{Goal 2:  Testing Methods on Different Components}} \\ \midrule
RQ 2.1 & How do existing work design testing methods for DL frameworks? \\
RQ 2.2 & How do existing work design testing methods for DL compilers? \\
RQ 2.3 & How do existing work design testing methods for DL hardware libraries? \\ \midrule
\multicolumn{2}{c}{\textbf{Goal 3:  Comparison Between Testing Methods}} \\ \midrule
RQ 3.1 & What are the strengths and weaknesses of existing DL framework testing methods? \\
RQ 3.2 & What are the strengths and weaknesses of existing DL compiler testing methods? \\
RQ 3.3 & What are the strengths and weaknesses of existing DL hardware library testing methods? \\ \midrule
\multicolumn{2}{c}{\textbf{Goal 4:  Exploration and Discussion}} \\ \midrule
RQ 4.1 & What are the findings from our survey? \\
RQ 4.2 & What are the challenges in existing DL library testing?\\ \bottomrule
\end{tabular}
\end{table*}

\subsection{Planning}
\label{sec:sub_plan}

\updd{Response to R2Q2: }{
To plan the review process, it is necessary to define our research goals and extract the research questions.
Our research goals are as follows.
\begin{enumerate}
    \item Goal 1: Collect state-of-the-art DL library testing methods and analyze the characteristics of DL library testing.
    \item Goal 2: Categorize the testing of DL libraries based on distinct components within the libraries and provide a mapping of the existing testing methods based on the various software testing techniques employed.
    \item Goal 3: Analyze and identify the strengths, weaknesses, and limitations of these different testing methods.
    \item Goal 4: Explore practical implications and findings in real-world scenarios and identify the challenges faced by existing testing methods, thereby delineating potential avenues for future investigation.
\end{enumerate}
}

\updd{Response to R2Q2: }{
We then further developed a set of Review Questions (RQs) for each of the defined goals to analyze three aspects in-depth.
The RQs are reported in~\autoref{tab:rqs}.
}






\subsection{Conducting}
\label{sec:sub_conduct}

\begin{figure}
    \centering     
    \includegraphics[width=0.9\linewidth]{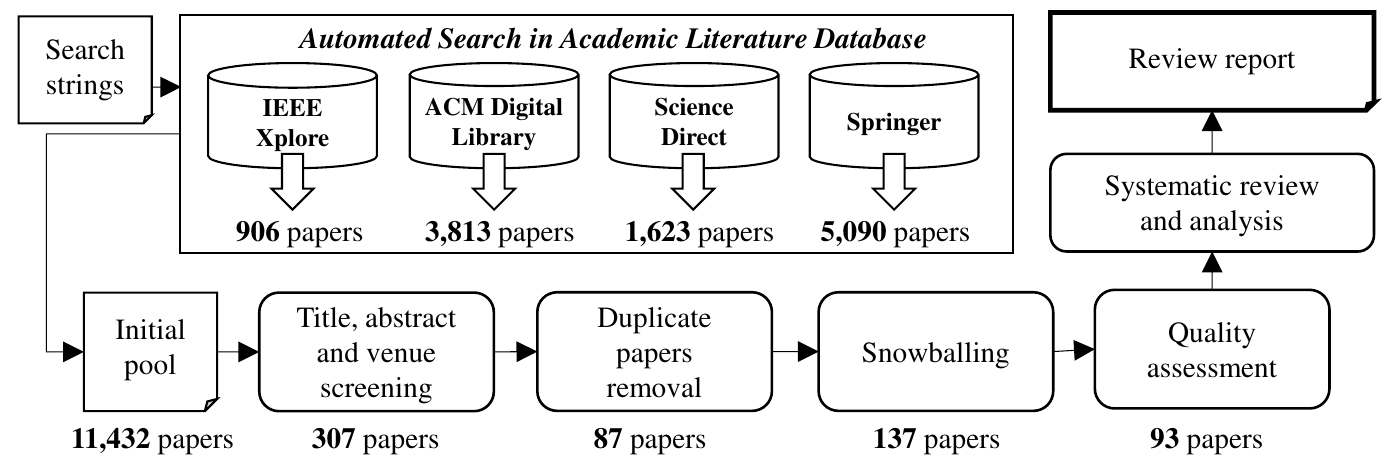}
    \caption{Process of Literature Selection}\label{fig:conductionpipe}
\end{figure}

\updd{Response to R2Q2: }{
This section introduces the research methods of this paper.
We first constructed a pipeline to search for relevant literature, ensuring comprehensive coverage of advanced related literature.
Subsequently, we conducted a step-by-step selection and quality assessment process, culminating in the formation of a final pool of papers.
To conduct the review, we took the following steps to search and collect the relevant papers.
The overall process is shown in~\autoref{fig:conductionpipe}.
}

\noindent
\(\bullet\)
\updd{Response to R2Q2: }{
Search String: To encompass the pivotal themes of our review, we meticulously crafted a search string that integrates keywords pertinent to deep learning libraries, compilers, frameworks, and associated bugs. This string is articulated in a syntax compatible with common search engines and databases, ensuring comprehensive coverage of relevant literature. The refined search string is delineated as follows:
}
\begin{lstlisting}[
    basicstyle=\small\ttfamily,
    breaklines=true,
    breakatwhitespace=true,
    frame=single,
    numbers=none,
    ]
(("Test" OR "Testing" OR "Fuzzing") AND ("DL Library" OR "Deep Learning Library" OR "DL Compiler" OR "Deep Learning Compiler" OR "DL Framework" OR "Deep Learning Framework" OR "DL Operator" OR "Deep Learning Operator")) OR "Deep Learning Library Bugs" 
\end{lstlisting}
\updmn{Response to R2Q2: }{
The search string consists of two major components connected by an `OR' operator.
The first component focuses on testing techniques for DL libraries, which is the primary scope of our survey. We use `Test OR Testing' as primary keywords since they effectively capture various testing methodologies, including metamorphic testing, mutation testing, and differential testing.
We specifically add `Fuzzing' as an additional keyword because some fuzzing-specific works might not be explicitly labeled with general testing terms.
Then we use `AND' to connect it with keywords such as `DL Library' and `DL Compiler' to ensure that these testing methods are relevant to DL library testing.
The second component, `Deep Learning Library Bugs', ensures the inclusion of broader empirical studies and bug analysis papers that might not explicitly mention testing methodologies.
Through this carefully designed search strategy, we ensure comprehensive coverage of both testing techniques specific to DL libraries and relevant empirical studies, thereby establishing a solid foundation for our systematic literature review.
}


\begin{wrapfigure}{r}{0.4\textwidth}
    \begin{center}
    \includegraphics[width=0.4\textwidth]{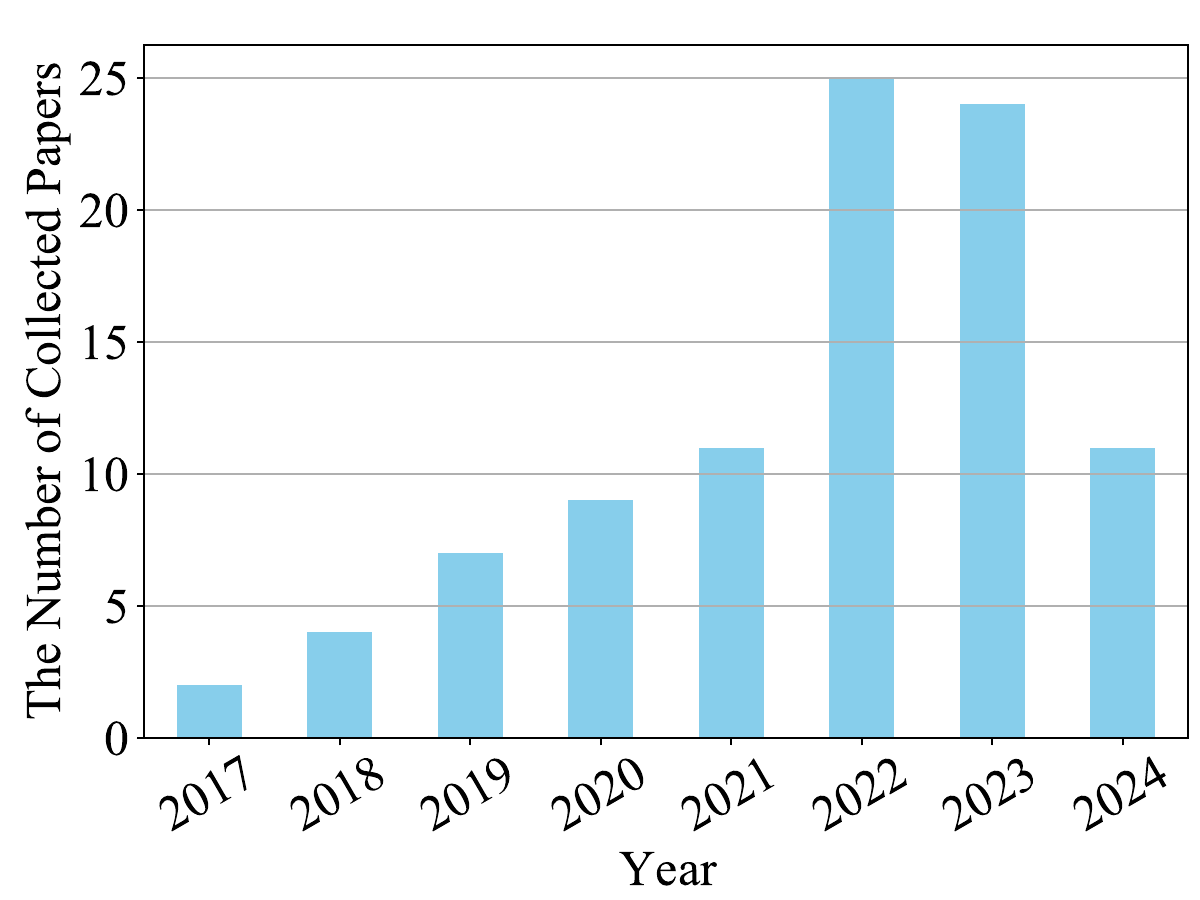}
    \caption{Number of Relevant Papers Per Year.}\label{fig:pub_year}
    \end{center}
\end{wrapfigure}
\noindent
\(\bullet\)
\updd{Response to R2Q2: }{
Search Scope: With our search string, we conducted an automated search across four widely recognized databases: IEEE Xplore, ACM Digital Library, Elsevier Science Direct, and Springer.
These databases were selected for their comprehensive coverage of published and recent papers in the fields of software engineering and computer science.
We collected the papers whose publication date is before September 1, 2024, in search.
}
    
\noindent
\(\bullet\)
\updd{Response to R2Q2: }{
Title, Abstract, and Venue Screening: We conducted a meticulous evaluation of the titles and abstracts to identify papers that were unequivocally relevant to the specific themes of our review.
To uphold the quality and credibility of the selected literature, we prioritized papers published in prestigious and highly regarded conferences (e.g., ICSE, ASE, FSE, OOPSLA) and journals (e.g., TOSEM, TSE) within the software engineering and program languages domains.
}

\noindent
\(\bullet\)
\updd{Response to R2Q2: }{
Duplicate papers removal: We removed duplicated entries collected from different sources to ensure a clean and unique set of papers.
}

\noindent
\(\bullet\)
\updd{Response to R2Q2: }{
Snowballing: Building upon the papers already collected and screened, we employed a forward and backward snowballing approach. This strategy involved tracing the references cited by the initial set of papers (forward snowballing) as well as the papers that cited them (backward snowballing).
}

\noindent
\(\bullet\)
\updd{Response to R2Q2: }{
Quality assessment: To mitigate the potential biases introduced by low-quality studies and to guide readers in discerning the reliability of conclusions, we conducted a rigorous quality assessment of the included papers.
Specifically, we invited two co-authors with expertise in both the fields of Software Engineering (SE) and Artificial Intelligence (AI) to entirely read the papers and assess the relevance, clarity, validity, and significance of the included papers. 
For any discrepancies in assessment results, we invited a third co-author to moderate the discussion and solve the differences. 
This process aimed to ensure that only studies of high methodological quality were incorporated into our review.
}

\subsection{Final Pool of Papers}
\label{sec:finalpool}

\updd{Response to R2Q2: }{
Employing the above comprehensive search strategy, our initial automated query across multiple academic literature databases yielded a substantial corpus of 11,432 papers, with contributions from IEEE Xplore (906), ACM Digital Library (3,813), Elsevier Science Direct (1,623), and Springer (5,090). Through a meticulous screening process on titles, abstracts, and publication venues, we distilled this collection to 307 papers aligned with our review objectives.
By removing duplicates and utilizing snowballing techniques, 137 papers entered our quality assessment, and 93 of them passed the manual assessment and formed our final paper pool.
This carefully curated set of 93 papers is the robust and reliable foundation for our comprehensive review analysis.
\autoref{fig:pub_year} exhibits the distribution of papers per year in the final paper pool.
It reveals a consistent upward trend beginning in 2017, underscoring a burgeoning scholarly interest in the field and the growing recognition of the critical importance of testing DL libraries. 
The marked escalation in research activity not only reflects the rapid maturation of the field but also signals a heightened awareness among the scientific community of the imperative need for robust, reliable, and secure deep learning frameworks.}

\updd{Response to R2Q2: }{
In the following sections, we have conducted a review and analysis of DL library testing methods to answer the RQs in~\autoref{tab:rqs} based on the collected papers.
}

\section{DL Library Testing Characteristics}
\label{sec:rq1}

\subsection{RQ1.1: DL Library Testing Property}
\label{s:libtestproperty}

\begin{wrapfigure}{r}{0.4\textwidth}
    \begin{center}
    \includegraphics[width=0.4\textwidth]{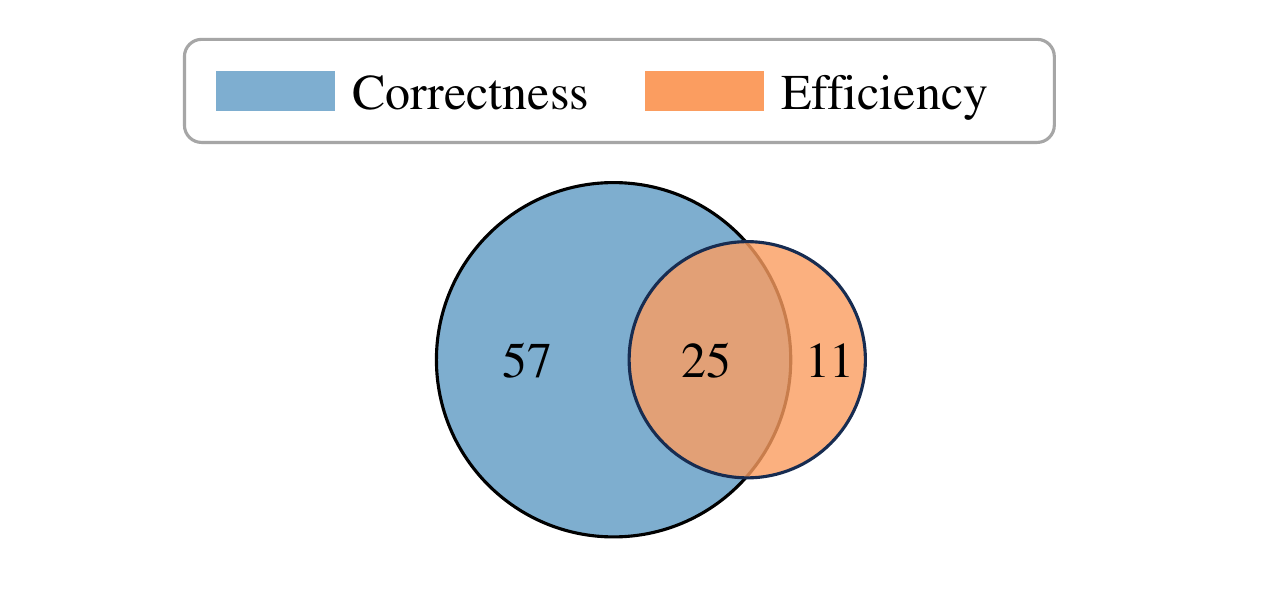}
    \caption{The Distribution of Test Properties involved in Collected Papers.}
    \label{fig:property}
    \end{center}
\end{wrapfigure}

The testing property refers to what the DL library testing methods test, which is directly related to the types of detected bugs in tests. 
It defines where the implementation of the DL library should meet the requirements and expectations.
\upd{Response to R1Q1 and R3Q13: }{
We analyzed 93 collected papers and found that existing DL library testing work mainly focuses on the functionality (\ie, \textbf{correctness}) and performance (\ie, \textbf{efficiency}) of the DL libraries.
They constructed test cases and conducted large-scale experiments to evaluate whether the functionalities of DL libraries are implemented correctly and whether the performance meets expectations.
\autoref{fig:property} uses a Venn diagram to show the number of papers involving different testing properties.}

\noindent
{\bf Correctness} measures the ability of a DL library to correctly perform its functions and complete the given task.
Correctness plays a vital role in the application and deployment of DL systems, which ensures the usability and trustworthiness of DL libraries.
When the correctness of the DL library is compromised, the intended function cannot be executed, which may cause three types of bugs, namely \textit{status bug}, \textit{numerical bug} and \textit{optimization bug}~\cite{chen2022toward,shen2021comprehensive}, which could be further exploited to endanger the safety and security of the whole DL system.
The status bug refers to the unexpected termination of valid inputs or illegal execution of invalid inputs on the DL library.
It includes various crashes, segmentation faults,
exceptions, etc.
The numerical bug and optimization bug occur when the DL library has incorrect behavior on valid inputs but does not crash.
At this time, the DL library will output wrong results and further affect subsequent calculations.
The former mainly consists of inconsistent outputs (\ie, inconsistency between expected and actual result) and NaN outputs (\ie, Not A Number, which is caused by overflow in the backend).
The latter happens when the DL library (especially the DL compiler) gets erroneous results or middle results in the optimization process, resulting in inequalities between before and after optimization.
Existing testing methods mainly test and validate the correctness of DL frameworks and compilers~\cite{pham2019cradle,deng2022fuzzing,xiao2022metamorphic}.

\noindent
{\bf Efficiency} evaluates the overhead of time, GPU memory, and other performance indicators of the DL library in executing a given task.
It determines the cost of large-scale deployment of DL libraries, which is of great significance for DL systems from the perspectives of performance, economics, and the environment.
The problem in DL library efficiency leads to the \textit{performance bug}.
This kind of bug not only severely compromises the usability of DL libraries, leading to less responsiveness and waste of computational resources,
but also puts pressure on both execution costs and the environment and results in a high carbon footprint~\cite{nistor2015caramel,jin2012understanding,patterson2021carbon}.
However, due to the limitations of testing methods and test oracles, existing work has paid limited attention to such bugs~\cite{guo2019empirical,levental2020comparing,zhang2024citadel}.
\upd{Response to R3Q13: }{
Among the 93 papers collected in~\autoref{sec:finalpool}, only 38.71\% involved performance bugs.
}

\subsection{RQ1.2: DL Library Testing Technique}

\begin{wrapfigure}{r}{0.4\textwidth}
    \begin{center}
    \includegraphics[width=0.4\textwidth]{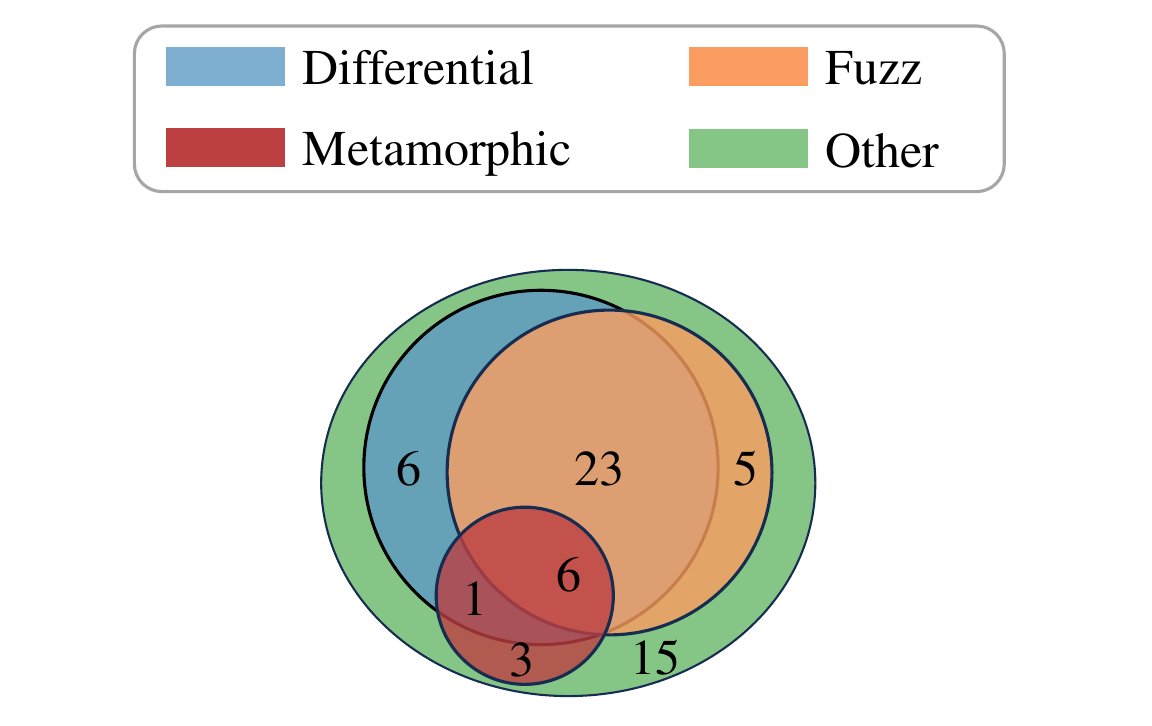}
    \caption{The Distribution of Test Techniques involved in Collected Papers.}
    \label{fig:technique}
    \end{center}
\end{wrapfigure}

The testing technique determines how the DL library testing methods test.
To effectively test different properties and identify bugs in DL libraries, 
researchers have employed traditional software testing techniques to design a variety of DL library testing methods.
Our study of collected papers indicates that the most widely spread techniques include differential testing, fuzz testing, and metamorphic testing.
\autoref{fig:technique} uses a Venn diagram to show the number of papers using differential testing, fuzz testing, metamorphic testing, and other test and evaluation techniques.

\noindent
{\bf Differential testing} is one of the most classic testing techniques in the field of SE, which usually processes the same input on two or more comparable implementations of a given software, uses the outputs between each other as the pseudo test oracle, and utilizes the difference between outputs to reveal potential bugs~\cite{mckeeman1998differential}.
\upd{Response to R3Q13: }{
As a simple but effective test technique, differential testing has not only achieved excellent results in traditional software testing and verification tasks~\cite{groce2007randomized,wang2021qdiff} but also shined in DL library testing~\cite{pham2019cradle,deng2022fuzzing} (36 of 93 collected methods used).
}
Different implementations of the same operator in different DL libraries or on different devices greatly facilitate the application of differential testing in DL library testing.

\noindent
{\bf Fuzz testing} is a classic software testing technique, which is widely used to automatically detect bugs such as crashes~\cite{liang2018fuzz, manes2019art}.
Fuzz testing typically generates a large number of test inputs and observes whether the target software or system can obtain expected outputs~\cite{liu2012software}.
Therefore, it is usually used to generate valid/invalid test cases in DL library testing, which is currently one of the most popular and effective techniques (34 of 93 collected methods used).
According to the test input generation methods, fuzzing can mainly be divided into generation-based fuzzing and mutation-based fuzzing~\cite{oehlert2005violating}.
The former generates test cases and inputs from scratch based on constraints or randomly, while the latter mainly mutates existing inputs to test more potential behaviors of DL libraries.
Since fuzz testing can generate a large number of inputs and achieve high API or code coverage in tests, testing methods based on it can often achieve outstanding results in real-world bug detection~\cite{xie2022docter,deng2023fuzzing}.

\noindent
{\bf Metamorphic testing} was proposed by Chen \et~\cite{chen1998metamorphic} in 1998 to construct test oracle.
Metamorphic testing constructs a series of metamorphic relations (MRs) that are from the necessary properties of the program under test.
An MR describes the expected change in the outputs of a target program when the inputs are changed.
Metamorphic testing can generate a series of test samples and judge whether the functionality of the program is as expected by comparing whether their results conform to the metamorphic relationship~\cite{segura2016survey}.
Researchers have designed some DL library testing methods based on the metamorphic testing technique to verify the computation and optimization of DL libraries, and successfully detected real-world bugs in the DL framework and compiler~\cite{wei2022free,xiao2022metamorphic} (10 of collected papers).

\subsection{RQ1.3: Characteristics of Different DL Library Components Tests}
\label{sec:rq13}

DL library component mentioned in~\autoref{sec:librarydefinition} points out the application and test objects of various DL library testing methods.
\upd{Response to R1Q4: }{In this section, we generally introduce the characteristics of the testing methods on the three DL library components and the differences between them.}
\upd{Response tot R3Q4: }{
Note that some libraries (\eg, TensorFlow and PyTorch) in the DL workflow have implemented general functions and APIs that can develop other ML algorithms and models.
Most existing testing methods~\cite{deng2023large,xie2022docter} construct test inputs for the various API of the entire DL library to discover bugs.
However, some testing methods~\cite{pham2019cradle,guo2020audee} rely on DL models as test inputs and only cover the library functions called in these DL models.
We have provided a detailed introduction and comparison of various testing methods in~\autoref{sec:framemethods}.
}

\noindent
{\bf DL framework testing} aims to discover bugs in DL frameworks, which can directly affect the output results and quality of DL models, leading to erroneous results or even crashes of DL systems built on these frameworks.
We collected and studied 69 papers on DL framework testing and observed that DL library testing has the following characteristics.
\ding{182} DL framework testing research has the most diverse testing methods and is the most well-developed field in DL library testing.
74.19\% of the collected papers in~\autoref{sec:finalpool} studied and tested DL framework bugs.
Existing research~\cite{islam2019comprehensive,yang2022comprehensive,chen2022toward,jia2021symptoms} has comprehensively investigated the characteristics, symptoms, and root causes of DL frameworks bugs through open-source communities (\eg, GitHub and Stack Overflow), providing valuable insights and guidelines for designing testing methods.
On this basis, researchers proposed a variety of testing methods and tools based on differential testing, fuzz testing, etc. to discover and identify DL framework bugs~\cite{pham2019cradle,deng2022fuzzing}.
\ding{183} DL framework testing methods accept manually built or automatically generated DL programs as input.
These test cases can be conveniently constructed using the user-friendly API provided by the DL frameworks.
\ding{184} The DL framework testing covers most of the bug types, including status bugs (\eg, crash, segmentation fault), numerical bugs (\eg, inconsistent output, NaN value), and performance bugs (\eg, unexpected time overhead).
Early framework testing methods mainly focused on several bugs in the forward and backward propagation stages of DL models~\cite{guo2020audee}.
With the development of testing methods, advanced testing methods~\cite{deng2023large} can cover thousands of APIs of the DL frameworks and discover dozens and even hundreds of bugs, which effectively promotes the development of DL frameworks.



\begin{wrapfigure}{r}{0.35\textwidth}
    \begin{center}
    \includegraphics[width=0.35\textwidth]{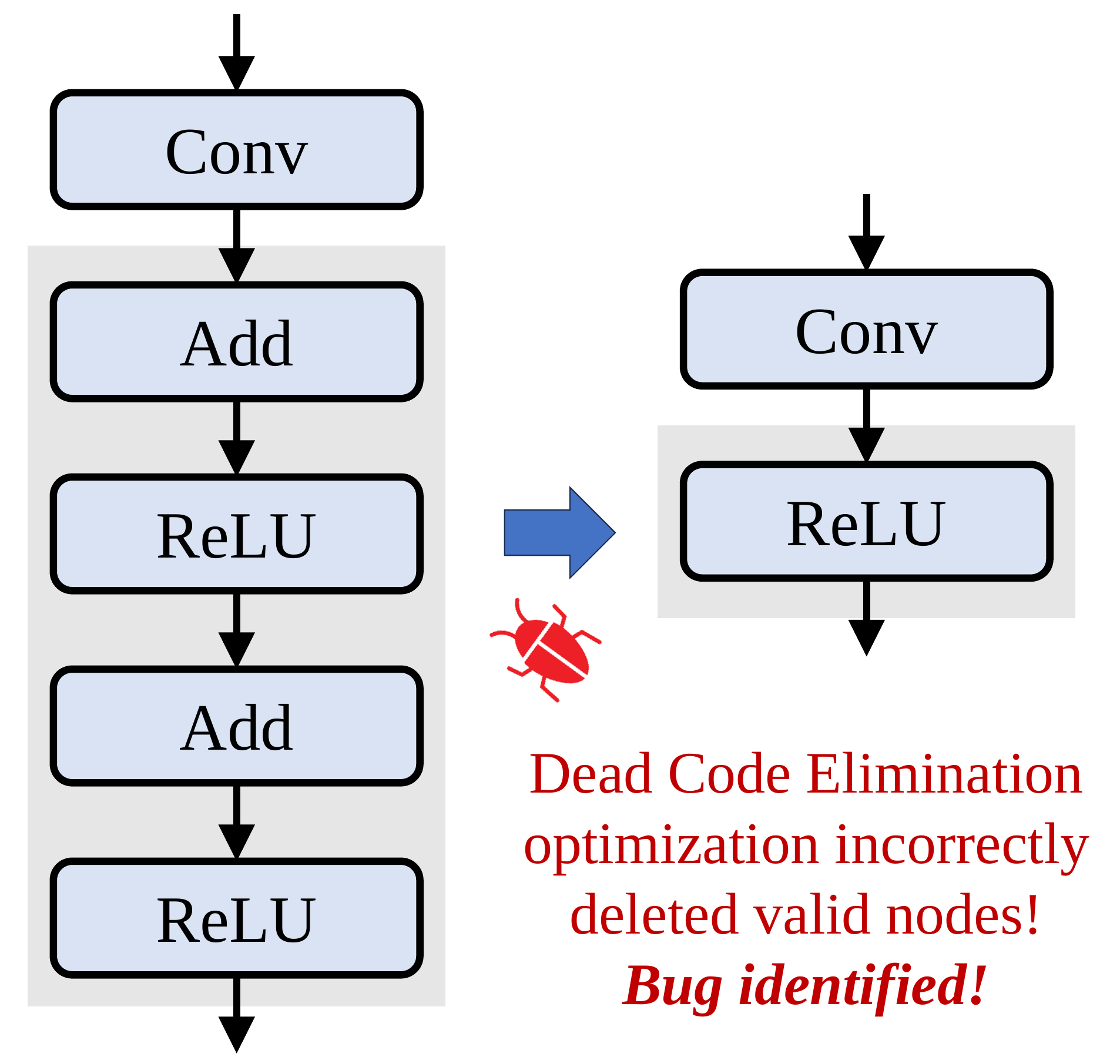}
    \caption{An Example Optimization Bug~\cite{xiao2022metamorphic} on the Glow Compiler.}
    \label{fig:mtdlcomp_demo}
    \end{center}
\end{wrapfigure}
\noindent
{\bf DL compiler testing} aims to find DL compiler bugs that cause the DL compiler to generate incorrect code, resulting in unexpected model behaviors~\cite{shen2021comprehensive}.
In traditional software, the compiler converts high-level program language (\eg, C++) to low-lever program language (\eg, assembly language) to create and optimize an executable program.
SE Researchers have leveraged techniques such as fuzzing~\cite{wu2023jitfuzz}, metamorphic testing~\cite{tao2010automatic}, differential testing~\cite{zhong2022enriching} and machine learning~\cite{chen2022boosting,chen2023compiler} to design a variety of testing methods for these traditional compilers (\eg, GCC, LLVM).
Different from traditional compilers, DL compilers convert abstracted DL models into optimized operators and code, which facilitates DL hardware libraries to perform calculations efficiently.
We have summarized the following characteristics from the collected DL compiler testing papers.
\ding{182} Similar to DL framework testing, DL compiler testing has also employed techniques such as fuzz testing and mutation testing. However, compared to DL framework testing, the diversity of methods and research in DL compiler testing is less extensive (only collected 12 papers in~\autoref{sec:finalpool}).
\ding{183} The DL compiler testing builds models and computation graphs as test input. Unlike DL framework testing that only focuses on parameters and tensor shape in test case generation, DL compiler testing also needs to meet constraints like semantic specification, posing challenges to efficiently build valid testing cases.
\ding{184} 
Due to the special architecture of DL compilers, DL compiler testing focuses on three bug-prone stages, namely model loading, high-level IR transformation, and low-level LR transformation stages.
The former loads an abstract DL model and transforms it into a computation graph, and the latter two respectively implement optimizations on high-level and low-level IRs.
Since specific conversion and transformation are involved, bugs are more likely to occur in the latter two stages~\cite{shen2021comprehensive}, therefore, the DL compiler testing methods often pay more attention to optimization bugs in these two stages.
Optimization bugs can cause semantic changes and inequality after the compilation process and ultimately incorrect calculation results or middle results~\cite{shen2021comprehensive}.
\autoref{fig:mtdlcomp_demo} shows an optimization bug in the Glow compiler, which incorrectly deletes layers (marked in the grey box) during the Dead Code Elimination (DCE) optimization, resulting in optimized operators outputting unexpected results. 

\noindent
{\bf DL hardware library testing} discovers bugs that can cause the DL model to obtain wrong results and abnormal runtime overhead in the calculation, which is difficult to perceive when executing the DL program or training a DL model.
There is relatively little research on DL hardware library testing (13 collected papers).
They mainly study and evaluate the functional correctness of DL hardware libraries~\cite{wang2020accuracy,uezono2022achieving}.
Since the DL hardware library is at the bottom of the entire DL workflow, as shown in~\autoref{fig:overview}, it is difficult to generate valid test code to detect potential bugs for these libraries, and little work is done to detect real-world bugs.

\noindent
\upd{Response to R1Q4: }{{\bf Summary: }
DL library testing for different components exhibits significant differences.
In terms of testing methods, DL framework testing employs diverse software testing techniques, including differential testing, fuzz testing, etc.
DL hardware testing, in contrast, primarily focuses on functional verification, with fewer related tests.
In terms of testing inputs, DL framework testing conveniently uses framework APIs to construct DL programs as inputs, while the testing inputs of DL compiler and hardware library testing need to satisfy additional optimization requirements and specifications, posing challenges in effective test case generation.
Finally, in terms of detected bugs, DL framework testing covers a variety of status, numerical, and performance bugs in the execution and computation of framework APIs, while DL compilers focus on optimization bugs during the optimization phase, which can lead to incorrect optimization, additional computational overhead, and erroneous results.
DL hardware library testing is relatively limited, primarily focusing on the correctness of DL hardware library implementations.
}

\section{DL Framework Testing}
\label{sec:framework}

\upd{Response to R3Q6: }{In this section, we first introduce the existing empirical studies on bugs and testing of DL frameworks and present DL framework testing methods according to different testing techniques used (RQ2.1).
We focus on introducing state-of-the-art and representative methods, which utilize specific testing techniques to break the research limitations at that time, and visually exhibit their descriptions and test results in~\autoref{tab:framwork_compare}.
The columns show the methods category, a brief description of each work, the test object, the number of bugs each work reported, and the type of detected bugs.
}
Then, we deeply compare and analyze the advantages and disadvantages of different testing methods and techniques (RQ3.1).
\upd{Response to R1Q6: }{
Additionally, we conduct experiments to compare five advanced open-source testing methods using different testing techniques in real-world scenarios (PyTorch and TensorFlow frameworks) to support our analysis of different methods.
}

\begin{table*}[]
    \caption{Representative DL Framework Testing Methods}
    \label{tab:framwork_compare}
    \centering
    \scriptsize
    \tabcolsep=5pt
    \begin{tabular}{crrrr}
    \toprule
    Category & \multicolumn{1}{c}{Method Description} & \multicolumn{1}{c}{Test Object} & \multicolumn{1}{c}{\# Bugs} & \multicolumn{1}{c}{Bug Type} \\
    \midrule
    \multirow{3}{*}{\begin{tabular}[c]{@{}c@{}}Empirical\\ Study\end{tabular}} & \begin{tabular}[c]{@{}r@{}}Summarized program bug reports and analyzed \\ the challenges of testing and localizing DL bugs~\cite{zhang2018empirical}\end{tabular} &  \multicolumn{1}{r}{TensorFlow} & / & / \\ \cmidrule{2-5}
    & \begin{tabular}[c]{@{}r@{}}Conducted large-scale study on DL frameworks \\ and summarized bug symptoms and root causes~\cite{chen2022toward}\end{tabular} &  \begin{tabular}[c]{@{}c@{}}TensorFlow/DL4J/\\ PyTorch/MXNet\end{tabular} & / & / \\ \midrule
    \multirow{8}{*}{\begin{tabular}[c]{@{}c@{}}Differential\\ Testing\end{tabular}} & \begin{tabular}[c]{@{}r@{}}Detected DL framework bugs via inconsistencies\\ between DL frameworks outputs~\cite{pham2019cradle}\end{tabular} & \begin{tabular}[c]{@{}r@{}}TensorFlow/\\ CNTK/Theano\end{tabular} & 12 & status/numerical\\ \cmidrule{2-5}
    & \begin{tabular}[c]{@{}r@{}}Mutated DL models to explore DL framework behaviors \\ and precisely localized the buggy layer in models~\cite{guo2020audee}\end{tabular} & \begin{tabular}[c]{@{}r@{}}TensorFlow/CNTK/\\Theano/PyTorch\end{tabular} & 26 & status/numerical\\ \cmidrule{2-5}
    & \begin{tabular}[c]{@{}r@{}} Extracted DL equivalence rules from documentation and open \\ source and constructed equivalent graphs to test~\cite{wang2022eagle}\end{tabular} & \begin{tabular}[c]{@{}r@{}}TensorFlow/\\ PyTorch\end{tabular} & 25 & status/numerical\\ \cmidrule{2-5}
    & \begin{tabular}[c]{@{}r@{}}Leveraged LLMs to generate test code and identified \\ bugs from the different results on different devices~\cite{deng2023fuzzing}\end{tabular} & \begin{tabular}[c]{@{}r@{}}TensorFlow\\PyTorch\end{tabular} & 65 & status/numerical\\ \midrule
    \multirow{8}{*}{\begin{tabular}[c]{@{}c@{}}Fuzz\\ Testing\end{tabular}} & \begin{tabular}[c]{@{}r@{}}Generate test cases for DL framework APIs and fuzz\\ DL framework based on open-sourced data~\cite{wei2022free}\end{tabular} & \begin{tabular}[c]{@{}r@{}}TensorFlow/\\ PyTorch\end{tabular} & 49 & \begin{tabular}[c]{@{}r@{}}status/numerical/\\ performance\end{tabular} \\ \cmidrule{2-5}
    & \begin{tabular}[c]{@{}r@{}}Extractd constraints from documentations to guide \\ test case generation and fuzz DL framework~\cite{xie2022docter}\end{tabular} & \begin{tabular}[c]{@{}r@{}}TensorFlow/\\ PyTorch\end{tabular} & 94 & status\\ \cmidrule{2-5}
    & \begin{tabular}[c]{@{}r@{}}Utilized the behavior of similar APIs as test\\ oracles and fuzz DL framework APIs~\cite{deng2022fuzzing}\end{tabular} &  \begin{tabular}[c]{@{}r@{}}TensorFlow/\\ PyTorch\end{tabular} & 162 & status/numerical\\ \cmidrule{2-5}
    & \begin{tabular}[c]{@{}r@{}}Designed AI semantics to compute and construct test \\ oracle and generate valid test cases in tests~\cite{schumi2022exais}\end{tabular} & \begin{tabular}[c]{@{}r@{}}TensorFlow\end{tabular} & 14 & status\\ \midrule
    \begin{tabular}[c]{@{}c@{}}Metamorphic\\ Testing\end{tabular} & \begin{tabular}[c]{@{}r@{}}Designed 11 metamorphic relations to verify the\\ correctness of DL framework functionality~\cite{ding2017validating}\end{tabular} & Caffe & / & status/numerical\\ \bottomrule
    \end{tabular}
\end{table*}

\subsection{RQ2.1: DL Framework Testing Methods}
\label{sec:framemethods}

\subsubsection{\upd{Response to R3Q7: }{Study on DL Framework Bugs}}
In the early stages of the development of DL framework testing, developers mainly relied on manual methods to identify and report bugs in DL programs~\cite{zhang2018empirical,zhang2019empirical}.
However, these early studies focused on analyzing programs and projects based on the DL framework and paid limited attention to the bugs in the DL framework itself.
To understand the symptoms and characteristics of various bugs, researchers~\cite{islam2019comprehensive} systematically studied over 3,000 bug posts and fixes related to five popular DL frameworks in the open source communities to understand the bug types, root causes, and their effects.
\upd{Response to R3Q3: }{
Although these studies have summarized the categories of DL framework bugs, they lacked an analysis of the symptoms and root causes of bugs.
To fill the gap, Jia \et~\cite{jia2020empirical} conducted empirical studies on the TensorFlow framework and found that the most common symptoms of DL framework bugs are functional errors and crashes, which received the most attention from following testing methods~\cite{deng2023fuzzing,xie2022docter}.
They also found that the most common root causes of bugs are errors related to data processing.
Jia \et~\cite{jia2021symptoms} further analyzed the fix patterns of DL framework bugs and identified 10 fix templates.
They have also studied multi-programming language (MPL) bugs in TensorFlow.
The code change complexity and communication complexity of MPL bug fixes were usually significantly higher than those of single programming language bug fixes, posing challenges for developers to locate and fix bugs~\cite{li2023understanding}.
Their findings have provided valuable insights and guidance for subsequent DL framework testing.
}

Recently, researchers have conducted more in-depth and subdivided research on DL frameworks and testing methods on the basis of prior work.
Some researchers deeply investigated and analyzed the DL framework bug and test oracles on different frameworks and further proposed new testing methods based on their findings~\cite{nejadgholi2019study,liu2020using,chen2022toward,kloberdanz2022deepstability,cui2022towards,jia2022injected}.
\upd{Response to R1Q3: }{Chen \et~\cite{chen2022toward} conducted a large-scale study on 1,000 bugs on 4 DL frameworks and found 13 types of root causes, including API misuse, numerical issues, etc.
Compared to prior studies, they provided a more comprehensive analysis of framework bugs and valuable insights into testing across different frameworks.
They further compared the indicators (\eg, line coverage) of the three existing test methods~\cite{pham2019cradle,guo2020audee,wang2020deep} to evaluate the effect of different testing methods and proposed a preliminary mutation-based testing tools TenFuzz.
They took the first step to evaluate the effectiveness of existing DL library testing methods.
However, they failed to evaluate more state-of-the-art open-sourced testing tools that cover more framework APIs in tests.}
The studies of Yang \et~\cite{yang2022comprehensive} and Harzevili \et~\cite{harzevili2022characterizing} further included the analysis of DL framework fixing patches.
\upd{Response to R3Q3: }{
Recently, researchers conducted studies on more subdivided bug types~\cite{tambon2024silent,hong2024investigating,velez2022challenges,liu2022taxonomy,du2022empirical,velez2022challenges}.
Tambon \et~\cite{tambon2024silent} first focused on silent bugs in Keras and TensorFlow frameworks.
Such bugs do not cause crashes or raise exceptions but do affect the DL framework's computation results and performance.
They systematically studied the symptoms and root causes of silent bugs and established four levels of impact for silent bugs, with the highest level affecting model output results.
Based on their finding, they called for the construction of unit test examples to diagnose silent bugs, providing valuable insights for testing and evaluation of related issues.
To fill the research gap of study on performance bugs, Cao \et~\cite{cao2022understanding} systematically studied the performance problems in DL frameworks such as TensorFlow and Keras and summarized five types of root causes from the aspects of API usage, model parameter selection, etc.
Based on the findings in the empirical study, they proposed and implemented a rule-based static checker, DeepPerf, to detect potential performance problems in DL systems.
}

Researchers have also paid attention to the DL framework bugs in subdivided deployment scenarios and environments such as distributed systems and JS systems~\cite{Chen2020comprehensive,quan2022towards,liu2023rise} and provided practical insights for DL framework deployment.
Aach \et~\cite{Aach2023large} focused on the distributed DL frameworks and studied the performance of ResNet models on PyTorch, Horovod, and DeepSpeed frameworks and different data loaders.
They found that using a suitable data loader can significantly accelerate the computation of ResNet models on these DL frameworks.
DeltaNN~\cite{louloudakis2023deltann} studied the impact of environmental factors such as DL frameworks on the performance of the image recognition model.
They have observed that the discrepancy in output labels of the same model on different DL frameworks can reach up to 72\% due to the noise introduced into the model weights by the conversion between the different frameworks.

\noindent
{\bf Summary and Analysis. }
\upd{Responsese to R3Q8: }{
Existing research leveraged interviews and empirical studies to summarize and analyze software bugs in DL frameworks and pointed out potential directions and challenges for designing testing methods for DL frameworks.}
\upd{Response to R1Q2: }{In the early stage, studies primarily focused on the symptoms and characteristics of bugs in individual popular DL frameworks like TensorFlow.
Recent research has devoted additional effort to a detailed analysis of the root causes of bugs on multiple DL frameworks or subdivided software problems such as silent bugs and has provided more detailed and timely observations and findings for tests than earlier studies.
Some researchers even further proposed new test tools based on their findings~\cite{harzevili2022characterizing,chen2022toward}.
Although researchers have a deep understanding of various DL framework bugs, there is still a lack of effective and diverse DL framework bug datasets or benchmarks to help developers evaluate and compare existing testing methods.}

\subsubsection{Differential Testing on DL Framework}
\label{s:frameworkdf}

To construct test oracles and identify DL framework bugs, researchers leveraged different implementations and devices to construct testing oracles based on the concept of differential testing.~\cite{pham2019cradle,guo2020audee,liu2018benchmarking,guo2019empirical,levental2020comparing}.
Depending on the generated test cases, the existing differential testing methods can be mainly divided into model-level testing and API-level testing~\cite{deng2022fuzzing}.

\begin{wrapfigure}{r}{0.4\textwidth}
    \begin{center}
    \includegraphics[width=0.4\textwidth]{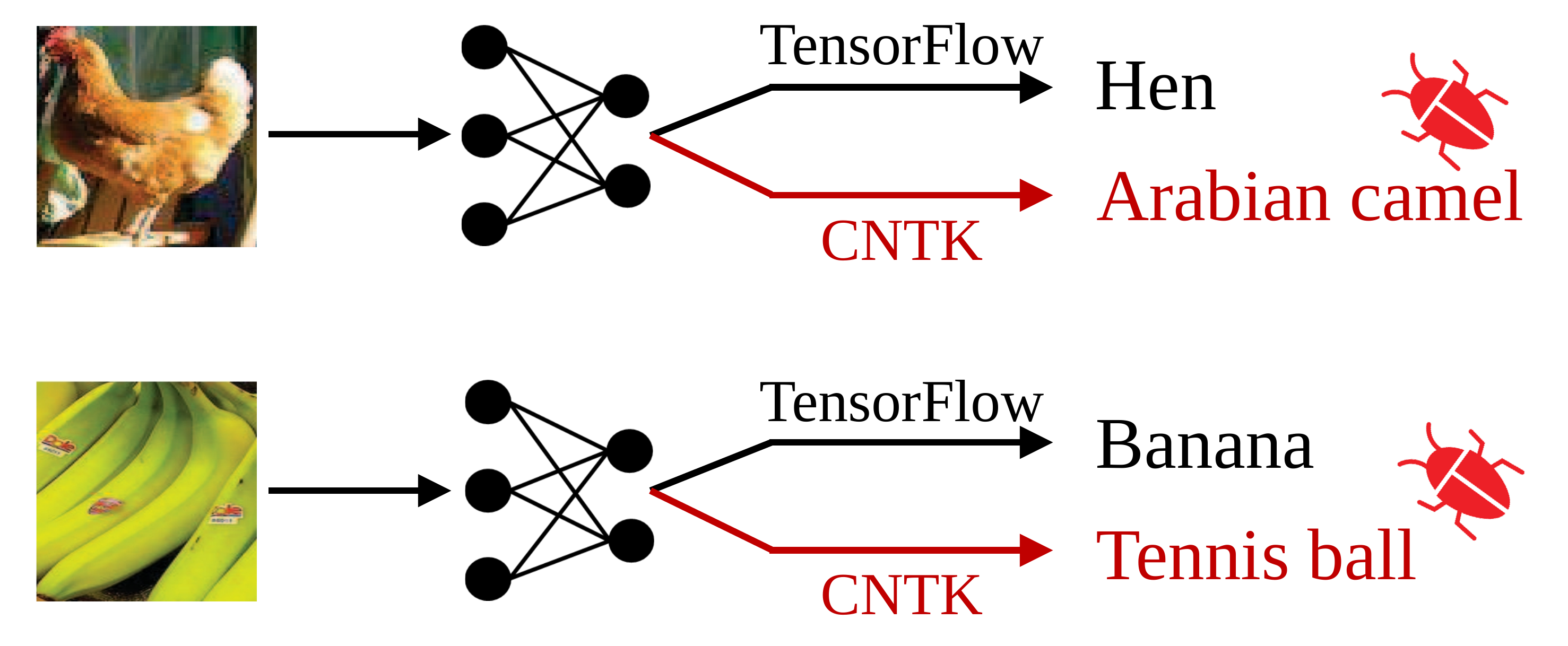}
    \caption{Trigger Inputs for Inconsistencies Between DL Frameworks}
    \label{fig:cradle_demo}
    \end{center}
\end{wrapfigure}

\noindent
{\bf Model-level Differential Testing. }
The model-level differential testing usually leverages the different results of a widely-used DL model (\eg, ResNet-50) on different platforms or frameworks to detect bugs~\cite{wang2020deep,li2022mmos}.
CRADLE~\cite{pham2019cradle} is one of the first tools to detect and identify bugs based on the concept of differential testing.
Based on Keras~\cite{ketkar2017introduction} which can build and train models on different DL frameworks as backends, CRADLE conducted differential testing on three frameworks(\ie, TensorFlow, CNTK, and Theano) and finally detected 12 bugs.
\autoref{fig:cradle_demo} shows two trigger figures of the inconsistencies that cause one model to have different prediction results and accuracy on different DL frameworks.
CRADLE compared the model layer outputs between multiple DL frameworks and detected such inconsistencies.
However, the inconsistent outputs of one layer may further lead to inconsistencies in subsequent layers, therefore this localization method is prone to false positives (FPs) and false negatives (FNs).
To break these limitations, Guo \et~\cite{guo2020audee} mutated the parameters of model layers to explore more DL model behaviors and leveraged a causal-testing-based technique to localize buggy layers and reduce the FPs.
\upd{Response to R1Q2: }{Although the above testing methods effectively detected bugs in popular DL frameworks like TensorFlow, Theano, and CNTK, they only focused on the model inference process and could not detect bugs in the backward propagation stage.
To fill the research gap, Gu \et~\cite{gu2022muffin} proposed Muffin, which creatively generated structure information and layer information of the DL model to thoroughly explore possible abnormal behaviors of the model.
Then it conducted differential testing between DL frameworks for model training and inference processes and finally detected 39 new bugs.
}
Although model-level differential testing methods obtain outstanding test results, they still have limitations in practice.
\upd{Response to R1Q2: }{The major limitation is that they can only explore limited DL model-related APIs and implementations in the framework
For example, existing research~\cite{wei2022free} indicated that LEMON~\cite{wang2020deep} only covered 35 TensorFlow APIs.
As a result, these methods cannot comprehensively test the entire framework (including APIs that can be used to develop ML algorithms), limiting the effectiveness of model-level differential testing methods.}

\noindent
{\bf API-level Differential Testing. }
Different from the DL model with dozens of layers generated by model-level methods, the test cases in the API-level method are simple calls or combinations of DL framework APIs.
Some test cases even merely call one operator or transformation APIs to calculate or process a set of randomly generated input data.
API-level testing eliminates the need to build and mutate DL models, which liberates the test cases from model shape constraints and enables the test to detect potential bugs on various framework APIs.
The existing API-level differential testing mainly used three approaches to construct test oracles.
\ding{182} In the early stage, researchers compared APIs implemented by different frameworks to identify bugs on DL frameworks such as TensorFlow and Theano~\cite{gu2022defect,prochnow2022diffwatch}.
However, these methods can hardly determine whether the different behaviors in tests come from implementation differences or real bugs in the DL frameworks, which limits the effectiveness of their testing.
\ding{183} Another simple but effective approach is to execute and compare DL APIs and operators on different devices (\eg, GPU and CPU) and observe the differences between the behaviors~\cite{zhang2021predoo,zhang2021duo,deng2023fuzzing,deng2023large}.
Zhang \et~\cite{zhang2021predoo} tested the precision errors by comparing the behaviors of seven DL framework operators on CPU and GPU.
However, their testing method can only construct test cases and explore abnormal behaviors for several DL APIs.
To efficiently and automatically generate test code at scale, Wei \et~\cite{wei2022free} proposed FreeFuzz, the first DL libraries testing method via mining from open source.
FreeFuzz first collected code that calls DL framework APIs from API documentation, DL framework test cases, and open-source models.
Then, it tracked and extracted the input and parameter constraints of each API from the execution of the collected code and constructed new test cases.
In the test, based on the concept of differential testing, FreeFuzz detected bugs by comparing the performance of test cases on different devices (\ie, CPU and GPU).
\ding{184} Differential testing using different frameworks and services only cover limited APIs that have different implementations on different frameworks and devices.
To break these limitations and detect more complex DL framework bugs, researchers tried to construct and leverage the equivalence relationship between APIs to design differential tests~\cite{yang2023fuzzing,deng2022fuzzing,xie2024cedar}.
Wang \et~\cite{wang2022eagle} proposed EAGLE, which created equivalent graphs to test DL frameworks.
They extracted 16 new DL equivalence rules from DL framework API documentation and non-crash issues in open-source communities and designed elaborate equivalent graphs that use different DL framework APIs, data types, or optimizations to produce identical output given the same input.
EAGLE focused on the numerical bug (\ie, inconsistencies) and finally detected 25 bugs on TensorFlow and PyTorch.
Deng \et~\cite{deng2022fuzzing} went one step further on the prior work and designed two elaborated equivalence (\ie, status equivalence and value equivalence) and matched thousands of API pairs based on these equivalence relations.
It considered the output values and status of APIs in a pair as test oracles for each other and effectively detected a total of 162 status and numerical bugs.

\noindent
{\bf Summary and Analysis. }
\upd{Response to R1Q2, R3Q4 and R3Q9: }{
In the early stage of DL framework differential testing, researchers usually compare the output results of DL models or APIs on multiple frameworks to detect potential bugs in the model layers (\eg, LEMON and Muffin).
However, these methods have two main limitations.
Firstly, they can only cover limited APIs in the framework, which need to be implemented on multiple frameworks and can be triggered by models or other test cases, which reduces the practical value of these testing methods.
For example, LEMON only covered 35 TensorFlow APIs~\cite{wei2022free}.
Secondly, they have a high false positive rate in tests.
Existing work~\cite{liu2023generation} has pointed out that the false positive rate of these model-level differential testing methods can reach 58\%.
Since test oracles rely on the implementation of multiple frameworks, it is difficult for developers to confirm whether the different results found in the tests are bugs or merely implementation differences, which affects the effectiveness of the test~\cite{guo2020audee, pham2019cradle}.
To effectively explore more potential bugs in DL frameworks, researchers use equivalent APIs or different devices to construct differential testing scenarios and achieve outstanding results with a much lower false positive rate.
However, some advanced methods still reported a high false positive rate of over 30\%~\cite{deng2023fuzzing}, which required a significant amount of manual effort to verify test results, increasing software maintenance costs.
How to reduce false positives while exploring various bugs is a challenge for differential testing research.
}

\subsubsection{Fuzz Testing on DL Framework}

According to the methods of generating test input, fuzzing can mainly be divided into generation-based fuzzing and mutation-based fuzzing~\cite{oehlert2005violating}.

\noindent
{\bf Generation-based Fuzz Testing. }
Generation-based fuzzing generates test inputs randomly or based on the specifications of test inputs.
As the test input of DL framework testing, DL programs usually have complex specifications (\eg, specific value ranges of API parameters, input sizes, and dimensions), and the input violating specification will lead to termination in execution, therefore existing work usually follow the specification to generate test inputs.
Existing testing proposed various methods to obtain specifications and constraints to guide test case generation~\cite{xie2022docter,kang2022skipfuzz,schumi2022exais}.
\ding{182} One direct approach is to directly extract and leverage API constraints from documentation or source code to guide testing~\cite{xie2022docter,shi2023acetest}.
Xie \et~\cite{xie2022docter} designed DocTer that analyzed documentation and extracted DL framework API constraints and further automatically built test cases.
The test case generation in DocTer generated both valid and invalid cases according to the constraints and specifications to comprehensively evaluate whether the DL framework has unexpected behaviors.
\upd{Response to R1Q2: }{DocTer provided a DL API constraint extraction method and an effective test input generation tool for DL framework testing, which facilitates and promotes the development of other methods~\cite{wang2022eagle}.
Unfortunately, only part of the DL framework APIs have detailed documentation that could provide constraint information, which limits DocTer's testing of some less commonly used APIs.
To overcome the limitations, Shi \et~\cite{shi2023acetest} proposed ACETest, which collected DL operators' information from the source code and extracted input validation constraints by analyzing the execution path, thus they could build valid test cases to uncover crashes in DL frameworks.}
\ding{183} Researchers also proposed the ML-based method to solve the constraints problem in tests~\cite{kang2022skipfuzz,liu2023neuri}.
SkipFuzz~\cite{kang2022skipfuzz} used active learning to learn the input constraints of different library APIs and generated valid test inputs for TensorFlow and PyTorch.
Deng \et~\cite{deng2023large} observed that prior methods typically generated valid inputs and were difficult to help tests cover edge DL library behaviors,.
To comprehensively explore the potential behaviors of DL frameworks, they designed zero-shot and few-shot learning to prime LLMs to generate edge cases while ensuring the semantic validity of generated test cases.
LLMs' knowledge of DL API calls enabled FuzzGPT to skip the step of collecting API constraints and directly generate effective test code for over 3,000 APIs and detect 108 bugs.

\noindent
{\bf Mutation-based Fuzz Testing. }
Mutation-based fuzzing usually applies various mutation strategies to introduce small changes to the valid test inputs, so as to explore potential bugs of DL frameworks while ensuring the validity of test cases as much as possible.
Some model-level testing methods and tools~\cite{guo2020audee,wang2020deep,shen2021deep,wu2022deepcov,nie2024python,zhang2021duo,chen2022toward,zhang2021predoo,10.1145/3691620.3695523,li2022mmos} mentioned above have implemented mutation operators to further explore potential behaviors in DL frameworks.
To increase the API coverage of the mutated DL models and explore DL framework behaviors, Zou \et~\cite{zou2023deep} leveraged a mutation-based hierarchical method to generate new models and effectively detected bugs on three frameworks.
They implemented two mutation modes, namely random mutation and heuristic mutation.
The former randomly modified the layers and operators of the model, and the latter was based on the results of the previously generated model and tended to select mutation operators that can increase the error of the model.


\begin{wrapfigure}{r}{0.4\textwidth}
    \begin{center}
    \includegraphics[width=0.4\textwidth]{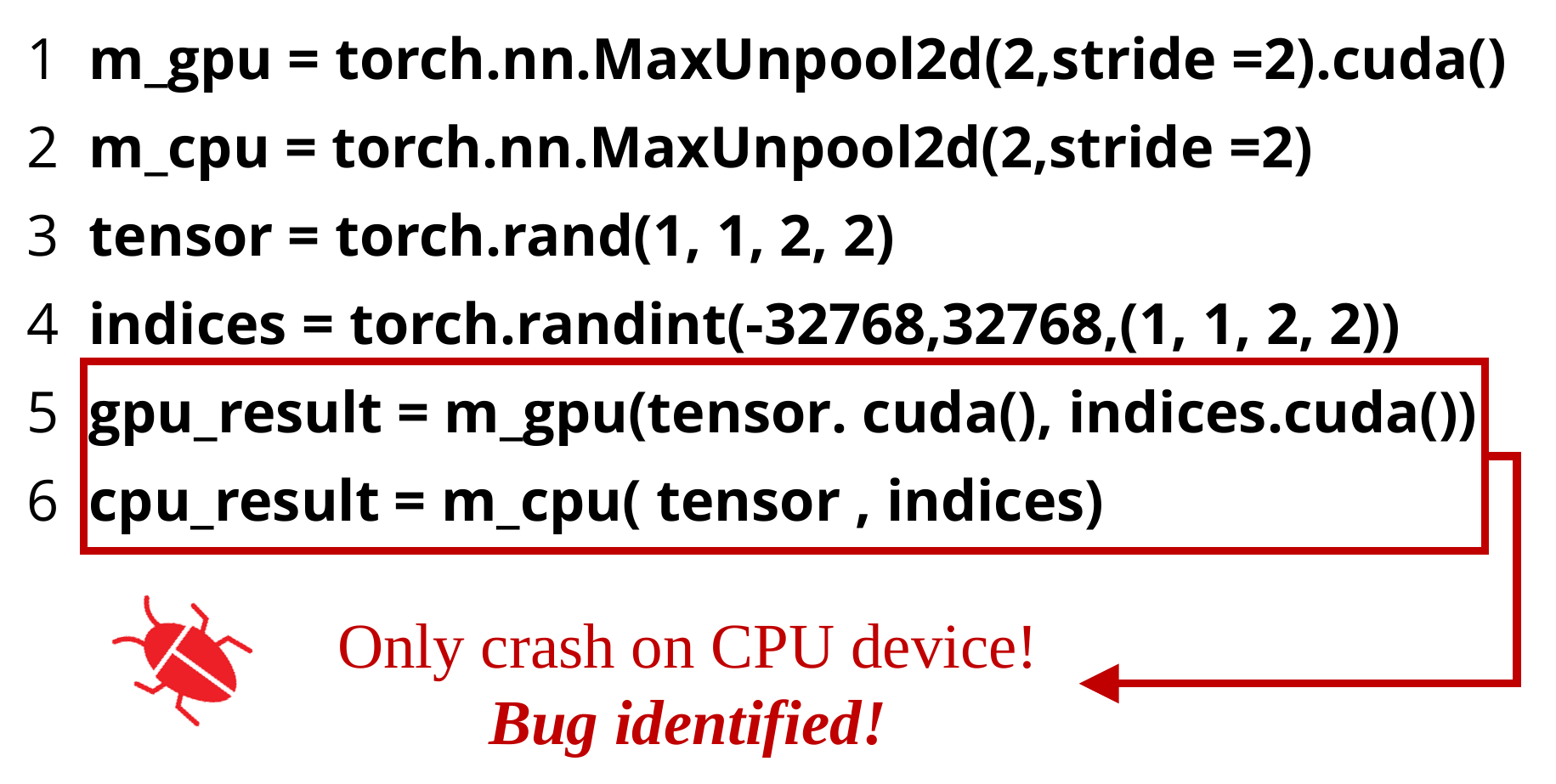}
    \caption{Invalid Input for \texttt{MaxUnpool2d} to Trigger Crash on CPU.}
    \label{fig:freefuzz_demo}
    \end{center}
\end{wrapfigure}
Researchers also proposed various API-level fuzz testing methods~\cite{wei2022free,li2023comet}.
FreeFuzz~\cite{wei2022free} collected code snippets from open source and implemented 3 categories of 15 mutation strategies on the data type and value to conduct fuzz testing.
The mutation strategies include mutating the dimension and datatype of a tensor, mutating the value of a tensor, etc.
Based on these strategies, FreeFuzz generated variants of a given test case from open source and further tested and revealed status, numerical and performance bugs for 1158 APIs on DL frameworks.
\autoref{fig:freefuzz_demo} shows a invalid input on \texttt{torch.nn.MaxUnpool2d} that only leads to a crash on the CPU\footnote{https://github.com/pytorch/pytorch/issues/68727}.
FreeFuzz generated the invalid input by mutating the value of the input tensor and identified this status bug.
Currently, developers have fixed this bug by adding a check for abnormal input values.
Researchers have implemented other mutation-based fuzz testing methods based on FreeFuzz~\cite{deng2022fuzzing,yang2023fuzzing,liao2024pso}.
However, limited by the design of mutation strategies, the above methods usually mutate a limited part of the test inputs (\eg, dimension) to generate valid mutated inputs, which cannot explore diverse code structures and potential framework behaviors.
To break the limitation, TitanFuzz~\cite{deng2023fuzzing} utilized the ability of LLMs in code generation to empower fuzz testing.
It leveraged a Codex model to generate test seeds for a given API and implemented four well-designed mutation operators.
These mutation operators masked the parameters of the API, the suffix and prefix of the test code, and the method under test in the seed, respectively, and then used the InCoder model to populate the mask and generate variants at scale.
TitanFuzz is currently one of advanced the testing tools, which covers a total of 3,544 APIs on PyTorch and TensorFlow frameworks in experiments.

\noindent
{\bf Summary and Analysis. }
\upd{Response to R1Q2 and R3Q9: }{
Fuzz methods in the early stage primarily utilize predefined rules and processes to extract constraints or perform mutations from code and documentation, exploring the behaviors of DL framework APIs (\eg, DocTer).
However, their test effects are determined by documentation information and mutation operators, limiting the exploration of the testing space and making it difficult to deeply uncover bugs in DL frameworks.
With the development of ML technologies, advanced ML techniques like LLMs have empowered fuzz testing, enabling tests to generate diverse test cases for thousands of DL framework samples, and exploring potential abnormal behaviors.
We have observed that ML-based testing methods (\eg, TitanFuzz) typically achieved better API coverage and test effects than predefined rule-based methods (\eg, DocTer), which have been demonstrated by our experiment results in~\autoref{fig:expframework}.
}

\subsubsection{Metamorphic Testing on DL Framework}

Metamorphic testing has been widely used to detect problems in DL models and systems~\cite{zhang2021deepbackground,arrieta2022multi}.
\upd{Response to R1Q2: }{Despite the success of metamorphic testing in DL testing, researchers have paid limited attention to the metamorphic testing of DL frameworks.
Ding \et~\cite{ding2017validating} constructed 11 MRS and validated AlexNet on the Caffe DL Framework.
Although they were the first to try to design MRs to validate a DL framework, they focused on the DL model accuracy in tests and paid less attention to DL framework bugs.
To address the limitation of localizing DL framework bugs at the parameter and tensor level, Chen \et~\cite{chen2024miss} proposed a testing framework with 18 MRs to test 10 common DL operators.
However, their manually designed MRs only cover popular DL operators like Conv2d and lack support for the thousands of other DL framework APIs.}

\updd{Response to R1Q2: }{
A significant challenge in metamorphic testing lies in the design of MRs.
MRs typically require expert knowledge and manual design, making it difficult to cover diverse DL framework APIs.
To improve the testing effect, researchers considered constructing MRs that apply to multiple APIs and combine them with other testing methods.
Liu \et~\cite{liu2023generation} designed 15 MRs to construct equivalent inputs for different DL frameworks and reduce the false positive rate in differential testing (\ie, 3.1\%).
Wei \et~\cite{wei2022free} combined fuzzing with metamorphic testing to detect performance bugs in DL frameworks by comparing the execution time of test cases on diverse APIs under the two data types of \texttt{float16} and \texttt{float32}.
The MR they designed is that programs carrying less precision information should execute faster.
It is undeniable that they have taken an important step in detecting performance bugs.
However, their MR can only provide a qualitative evaluation of API performance and cannot accurately and quantitatively detect and identify unexcepted runtime overhead.}

\noindent
{\bf Summary and Analysis. }
\upd{Response to R1Q2: }{
The major limitation of metamorphic testing methods is that the quality of MRs directly affects the results of metamorphic testing.
Simple MRs in early research (\eg, the impact of extending datasets~\cite{ding2017validating}) can hardly uncover bugs in DL frameworks and have limited test effects, while complex MRs (\eg, the impact of transposing the input and kernel of convolution operator~\cite{chen2024miss}) can verify the functionality of operators in-depth, it is difficult to cover diverse APIs.
To overcome these limitations, recent research has combined MRs with other testing techniques (\eg, fuzz testing) to provide test oracles for more framework APIs and comprehensively verify DL framework behaviors~\cite{wei2022free}.
}

\subsubsection{Other Testing on DL Framework}

In addition to the aforementioned testing techniques, researchers have used various software testing techniques such as smoke testing and just-in-time defect prediction to verify the functional correctness and efficiency of the DL frameworks~\cite{herbold2022smoke,zhang2024citadel,jia2021unit,ge2023justintime,wei2024demystifying}.
Their methods have provided valuable insights and new perspectives for DL framework testing.
However, these methods have rarely detected real-world bugs in DL frameworks, which limits their effectiveness.
\upd{Response to R3Q3: }{
Jia \et~\cite{jia2021unit} conducted mutation testing to validate the quality of the unit test cases of three DL frameworks.
Different from the aforementioned mutation-based fuzz testing, mutation testing observes whether unit test cases can identify the modified DL frameworks and kill them.
They designed 13 categories of mutation operations to mutate the DL framework to generate mutants.
Their experiment showed that more than 60\% mutants were not detected by the unit test cases in the framework, which points out the ineffectiveness of these test cases and possible research opportunities.}
To promote the development of the testing methods for DL frameworks and systems, Kim \et~\cite{kim2021denchmark} proposed an open-source DL bug dataset, covering 4,577 bugs in 8 categories of DL software, including DL frameworks, platforms, compilers, etc.
Their dataset mainly included bug reports and statistical information like buggy entities.
How to automatically reproduce these DL framework bugs on the corresponding environment to facilitate the evaluation of existing testing methods remains a challenge.

\subsection{RQ3.1: Comparision and Analysis}
\label{sec:framedwork_compare}

\begin{figure}
    \centering     
    \includegraphics[width=0.9\linewidth]{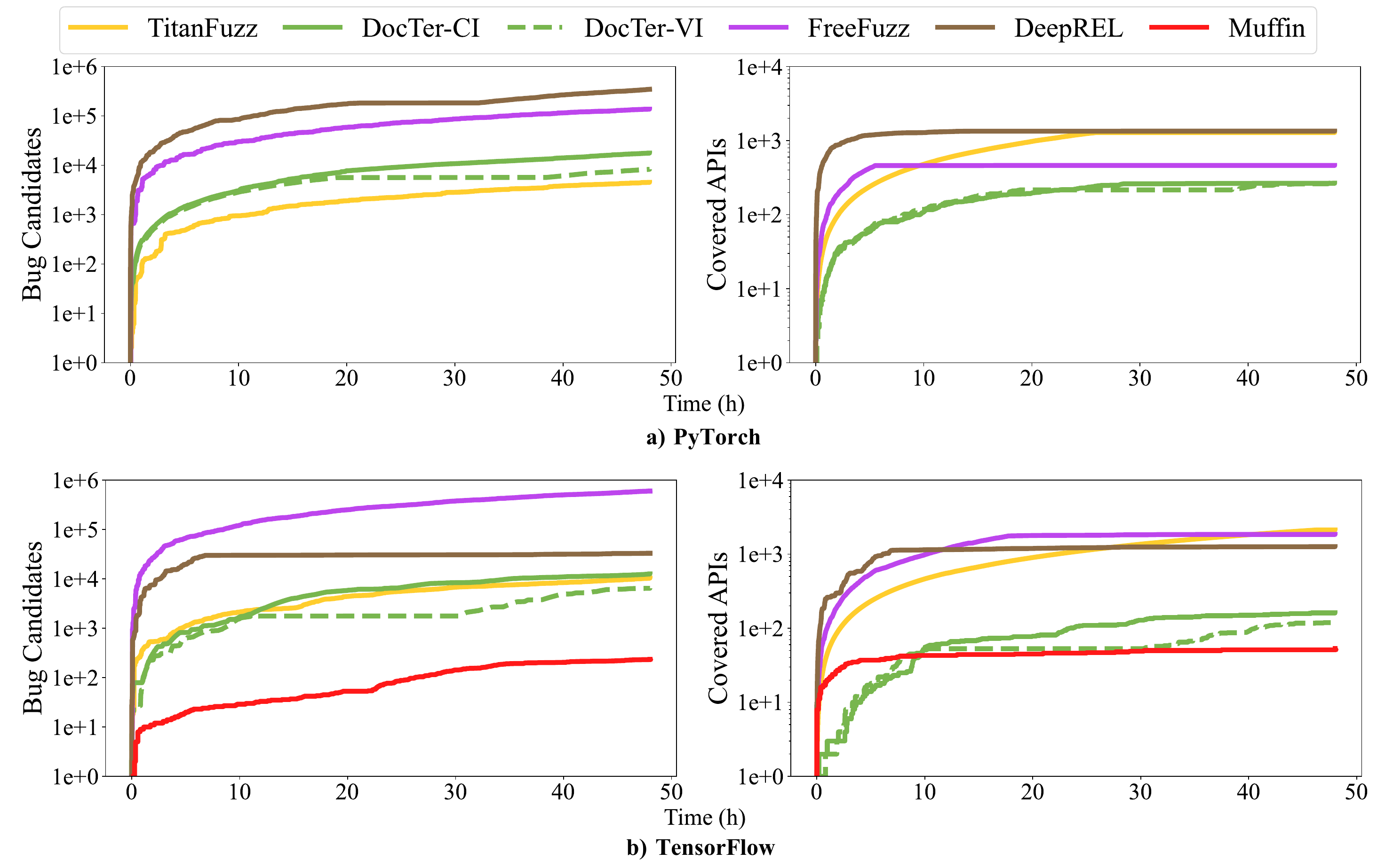}
    \caption{\updmn{Response to R3Q2: }{Comparison of Testing Methods on the PyTorch and TensorFlow Frameworks}}\label{fig:expframework}
    \vspace{-10pt}
\end{figure}

\upd{Response to R1Q2 and R1Q6: }{{\bf Comparison Experiment and Results. }
To compare the effectiveness of testing methods based on different testing techniques and principles, we conducted experiment with five state-of-the-art and representative open-source testing methods, covering three popular testing techniques and both model-level and API-level methods, namely Muffin~\cite{gu2022muffin}, FreeFuzz~\cite{wei2022free}, DocTer with conforming (CI) and violating inputs (VI)~\cite{xie2022docter}, DeepREL~\cite{deng2023fuzzing}, and TitanFuzz~\cite{deng2023large}.
We have followed the instructions in their source code and repeatedly executed their source code for 48 hours in the same environment to separately detect bugs on TensorFlow and PyTorch frameworks.
Note that the final outputs of existing testing methods were thousands of bug candidates, which requires dozens of man-months to validate these candidates, and how to validate and filter out invalid bug candidates automatically is out of our scope.
Therefore, we directly recorded \ding{182} the API coverage and \ding{183} the number of bug candidates generated by each method in experiments.
More details of the experiment results and settings are in our repository~\cite{ourrepo}.
\updmn{Response to R3Q2: }{The experimental results are shown in~\autoref{fig:expframework}, where the X-axis represents time and the Y-axis represents API coverage and the number of generated bug candidates in scientific notation.}
In experiments, Muffin covered the fewest APIs (52 TensorFlow APIs) and generated 233 bug candidates within 48 hours.
TitanFuzz achieved the largest API coverage with a total of 3,413 APIs on two frameworks.
Note that, FreeFuzz and DeepREL have generated a large number of bug candidates, with 737,070 and 379,478 for the two frameworks respectively, which is 49.67 times and 25.57 times the detected buggy candidates of TitanFuzz.
However, we have not found an efficient way to automatically verify these bug candidates.
So many bug candidates require a significant amount of manual verification and screening, which poses a challenge for comparing the testing effects of different testing methods in real-world scenarios.
}



\updd{Response to R1Q2 and R1Q6: }{{\bf Analysis. }
Based on the collected literature and experiment results in~\autoref{fig:expframework}, we analyzed the advantages and drawbacks of the testing methods based on different testing techniques and principles as follows:
}

\noindent
\(\bullet\)
\updd{Response to R1Q2 and R1Q6: }{{\it Differential Testing. }
Existing differential testing can provide general test oracles for various DL framework APIs (\eg, comparisons between different devices and different DL framework implementations), and effectively identify the status and numerical bugs.
However, existing differential testing methods have two major limitations.
\ding{182} Compared to metamorphic testing whose MRs are manually designed based on expected behaviors, the test oracles in differential testing are not accurate.
Similar APIs or different framework implementations used in differential testing may have inconsistent output results, which leads to false positives in testing (\eg, DeepREL's false positive rate exceeds 30\%), requiring manual effort to validate the buggy cases.
Such inconsistency also makes it difficult to detect performance bugs.
\ding{183} The test oracle design could affect the API coverage in differential testing.
Muffin, which utilized mutated models to conduct differential testing across multiple frameworks, can only cover APIs that are called by the models and have corresponding implementations in different frameworks, ultimately covering 52 APIs in our experiments.
In contrast, DeepREL, leveraging the status and output results of equivalent APIs to construct test oracles, effectively explored 2,607 APIs within 48 hours.
Such an observation of different is consistent with our analysis in~\autoref{s:frameworkdf}.
}

\noindent
\(\bullet\)
\updd{Response to R1Q2 and R1Q6: }{{\it Fuzz Testing. }
Existing fuzz testing methods have generated thousands of valid test inputs by extracting constraints, mutating seed inputs, etc., covering a variety of DL framework APIs and comprehensively exploring framework behaviors.
The development of ML techniques (\eg, LLMs) has further improved the effect of fuzz testing in test case generation.
LLM-guided fuzzing method TitanFuzz covered 3,413 APIs within 48 hours, the most among all methods in our experiments. 
Existing fuzz testing methods have two major limitations.
\ding{182} The effectiveness of fuzz testing is correlated with the way they generate test cases (\eg, document-based, ML-based).
DocTer extracted API constraints from documentation to generate test cases, therefore it can only test APIs with detailed documentation descriptions.
The two kinds of inputs in DocTer finally covered 470 APIs in 48 hours of testing.
In contrast, TitanFuzz leveraged LLM to generate and mutate test cases and covered 3,413 APIs in our experiments.
\ding{183} Fuzzing cannot provide test oracles in tests.
Merely using fuzz testing can only detect simple status bugs such as crashes~\cite{shi2023acetest}.
Therefore, existing fuzzing methods typically utilized differential testing techniques to construct test oracles, effectively identifying the status and numerical bugs~\cite{deng2022fuzzing,xie2022docter}.
Titanfuzz used differential testing to construct test oracles and effectively discovered buggy cases by identifying the difference between DL framework behaviors on GPU and CPU.
FreeFuzz, further combined with metamorphic testing, can further detect performance bugs.
}

\noindent
\(\bullet\)
\updd{Response to R1Q2 and R1Q6: }{{\it Metamorphic Testing. }
Compared with other techniques, existing DL framework research using metamorphic testing is relatively limited.
The MRs in metamorphic testing are abstracted from the expected functionalities of DL frameworks, providing accurate test oracles for identifying status, numerical, and performance bugs.
It still has two limitations.
\ding{182} The design of MRs requires the manual effort of experts with domain knowledge.
Compared to test oracles of differential testing that are general for most APIs (\eg, results on different devices), MRs are much more complex and typically designed for a specific operator (\eg, the impact of transporting the input and kernel of the specific operator).
\ding{183} Metamorphic testing methods can only verify whether the target framework is implemented as expected in aspects related to MRs, but cannot discover potential bugs in other aspects, making it difficult to explore various buggy behaviors of DL frameworks.
Therefore, current research (\eg, FreeFuzz) mainly combines it with other testing techniques to provide test oracles with fewer false positives~\cite{wei2022free}.
}

\section{DL Compiler Testing}
\label{sec:dependency}



Compared with DL frameworks, DL compilers require higher professional knowledge requirements and are often used to solve task-specific problems of the underlying optimization of DL models, which makes it difficult to design testing methods for DL compilers.
As a result, DL compiler testing has received less attention and are still in its infancy in recent years.
\upd{Response to R3Q6: }{Similar to~\autoref{sec:framework}, we introduce the existing testing methods for DL compilers according to the testing techniques used (RQ2.2), and then discuss the advantages and limitations of these methods (RQ3.2).
\autoref{tab:compiler_compare} present representative DL compiler testing methods using different testing techniques.
}
In addition, we conduct experiments on the TVM compiler, which is one of the most popular DL compilers, to compare and analyze the testing effects of three advanced open-source testing methods.

\begin{table*}[]
    \caption{Representative DL Compiler Testing Methods}
    \label{tab:compiler_compare}
    \centering
    \scriptsize
    \tabcolsep=5pt
    \begin{tabular}{crrrr}
    \toprule
    Category & \multicolumn{1}{c}{Method Description} & \multicolumn{1}{c}{Test Object} & \multicolumn{1}{c}{\#Bugs} & \multicolumn{1}{c}{Bug Type} \\
    \midrule
    \begin{tabular}[c]{@{}c@{}}Empirical\\ Study\end{tabular} & \begin{tabular}[c]{@{}r@{}}Studied and analyzed DL compiler bugs and\\ summarized the bug symptoms and root causes~\cite{shen2021comprehensive}\end{tabular} & \begin{tabular}[c]{@{}r@{}}TVM, Glow,\\ and nGraph\end{tabular} & /& / \\ \midrule
    \multirow{6}{*}{\begin{tabular}[c]{@{}c@{}}Fuzz\\ Testing\end{tabular}} & \begin{tabular}[c]{@{}r@{}}Generated valid test cases based on constraints\\ and conducted differential testing on DL compilers~\cite{liu2023nnsmith}\end{tabular} & \begin{tabular}[c]{@{}r@{}}TVM, ONNXRuntime,\\ TensorRT, and   PyTorch\end{tabular} & 72 & status/numerical \\ \cmidrule{2-5}
    & \begin{tabular}[c]{@{}r@{}}Mutated low-level IR of DL compilers and\\ conducted coverage guided fuzz testing\end{tabular}~\cite{liu2022coverage} & TVM & 49 & \begin{tabular}[c]{@{}r@{}}status/numerical/\\ performance\end{tabular} \\ \cmidrule{2-5}
    & \begin{tabular}[c]{@{}r@{}}Used domain-specific languages to describe \\ constraints and generate valid test inputs\end{tabular}~\cite{wang2023gencog} & TVM & 16 & \begin{tabular}[c]{@{}r@{}}status/numerical\end{tabular} \\ \midrule
    \end{tabular}
\end{table*}

\subsection{RQ2.2: DL Compiler Testing Methods}


\subsubsection{Study on DL Compiler Bugs}

\upd{Response to R1Q2: }{
Researchers have conducted empirical studies on DL compilers bugs, analyzing types and root causes of bugs on popular DL compilers, including TVM, Glow, PlaidML, etc~\cite{shen2021comprehensive,du2021empirical}.
They observed that crashes and wrong code (\ie, optimization bugs) are the most common bugs in the DL compiler and the IR transformation stages were the most buggy stages, which provided insights for DL compiler testing methods.
To fill the gap in prior studies on optimizing efficiency, Verma \et~\cite{verma2021performance,verma2022towards} focused on the optimization performance of DL compilers such as TensorFlow Lite and TensorRT (TF-TRT) in edge inference scenarios.
Their large-scale evaluation revealed that integrated TF-TRT consistently performed better at high-precision floating-point operations but could not hold at low precision.
Although the above work discussed the performance of DL compilers, they mainly focused on optimization performance, failing to delve into the potential bugs behind low performance or provide insights for detecting performance bugs.}

\noindent
{\bf Summary and Analysis. }
\upd{Response to R1Q2: }{
Existing empirical studies on DL compilers collect bug reports from the open-source community, analyze their symptoms and root causes, and further provide valuable suggestions for the development, application, and testing of DL compilers.
However, they still lack investigation and research on some DL compilers (\eg, XLA) and fix patterns of DL compiler bugs, which can promote the research on testing and repair of DL compiler bugs.
Furthermore, researchers mainly focus on compiler bugs related to functional correctness, such as crashes and optimization bugs, only Shen \et~\cite{shen2021comprehensive} have studied the bad performance and root causes of DL compilers.
}

\subsubsection{Fuzz Testing on DL Compiler}

As we have discussed in~\autoref{sec:rq13}, DL compiler testing faces the challenge of generating valid test cases and exploring the potential behaviors of DL compilers.
To overcome the challenge, researchers have leveraged the fuzz testing techniques to generate test cases based on generation and mutation.

\noindent
{\bf Generation-based Fuzz Testing. }
Researchers have proposed a series of generation methods that follow semantical constraints~\cite{ren2023effective,liu2023nnsmith}.
\upd{Response to R1Q2: }{
Liu \et~\cite{liu2023nnsmith} proposed NNSmith that leveraged the operator constraints provided by users to generate valid graphs for DL compiler testing and performed cross-validation on multiple DL compilers.
Recently, researchers have observed that previous methods, while generating test cases that conform to constraints, have difficulty carrying enough effective information (\eg, operators and attributes) to explore DL compiler behavior.
Moreover, they often require heavy manual effort to write constraints for each operator in tests.
To overcome these limitations, researchers proposed GenCoG which used domain-specific languages to describe the constraints of operators and automatically generate valid graphs based on feasible combinations of input tensor types and attributes to carry useful information~\cite{wang2023gencog}.}
It finally covered 62 operators and detected 14 bugs on TVM.
Unfortunately, their implementation only supported the TVM compiler.
Generalizing to various DL compilers can increase practical value.

\noindent
{\bf Mutation-based Fuzz Testing. }
In addition to generation-based fuzzing methods, researchers also pay attention to mutation-based fuzz testing~\cite{lin2023deepdiffer,ma2023fuzzing}.
\upd{Response to R1Q2: }{Based on the findings from their empirical study, Shen \et~\cite{shen2021comprehensive} designed the testing tool TVMfuzz for the TVM compiler.
It conducted fuzz testing based on the directed graph built by TVM APIs and mutated the shape and type of tensors to construct new unit tests.
It finally detected 8 crash bugs by performing differential testing on the two versions of TVM.
However, TVMFuzz can only detect obvious bugs such as crashes, but cannot catch numerical bugs.
To break through the limitation, DeepDiff~\cite{lin2023deepdiffer} implemented priority-guided fuzz testing that efficiently identifies logic errors by mutating IR seeds to maximize the difference between results on different TVM versions and detect 9 new bugs.
}
Liu \et~\cite{liu2022coverage} further proposed the first coverage-guided fuzz testing tool for testing the tensor compiler (\ie, TVM), Tzer.
They designed six kinds of mutation operators (\eg, inserting loops, replacing operations) for the low-level IR of DL compilers to trigger more potential behaviors, and added the mutated IR to the seed pool when it covered more code.
It finally detected 49 new bugs on the TVM compiler, surpassing prior tools (\eg, TVMfuzz) in terms of coverage and bug detection.
However, the above fuzzing methods mainly focus on the IR transformation stages which are related to optimization, and ignore the model loading stage.
To fill the research gap and cover more DL compiler operators, Shen \et~\cite{DBLP:journals/corr/abs-2407-16626} transferred the knowledge of DL framework fuzzers (\eg, DocTer and DeepRel) to generate effective test cases for diverse DL compiler operators to detect crashes and inconsistencies.

\noindent
{\bf Summary and Analysis. }
\upd{Response to R1Q2: }{
Fuzz testing can effectively generate test inputs for DL compilers and explore potential behaviors and has already helped developers identify hundreds of bugs on these compilers.
In the early stages, DL compiler fuzzing methods utilized simple random generation or mutation on tensor shapes to generate test cases~\cite{shen2021comprehensive,ren2023effective}.
With the advancement of technology, researchers have proposed a series of methods such as coverage-guided fuzzing and priority-guided fuzzing to efficiently generate valid test cases and deeply explore DL compiler behaviors.
These testing methods can cover more DL compiler operators and be combined with differential testing to detect more types of bugs.
Despite this, it can be observed that existing DL compiler fuzzing methods still rely on predefined rules and constraints to construct and mutate test cases.
Compared to state-of-the-art DL framework fuzzing methods using ML techniques~\cite{deng2023large,li2022alphaprog}, these DL compiler fuzzing methods still have a lot of room for improvement.
}

\subsubsection{Other Testing on DL Compiler}


The DL compiler fuzzers mainly construct test oracles by leveraging the concept of differential testing (\eg, results between different devices), enabling the detection of state and numerical bugs, but they cannot effectively identify complex optimizations or performance bugs.
To construct test oracles to discover more complex optimization bugs, researchers have delved into metamorphic testing.
Xiao \et~\cite{xiao2022metamorphic} designed elaborated metamorphic relations to test DL compilers.
For example, they introduced a series of operators whose final result is 0 in the input, and then in tests and then observed the change of the DL compiler result to detect potential bugs.
They have detected 4 DL compiler bugs on three compilers (\ie, TVM, Glow, and XLA), and \autoref{fig:mtdlcomp_demo} provides one example bug.
This bug causes the DCE optimization in Glow to mistakenly delete active and valid nodes in the computation graph, resulting in the generated optimized operators outputting unexpected results.
In addition, to alleviate the false positives introduced by differential testing on different compilers, Zhou \et~\cite{10.1145/3689757} proposed a method that builds tensor graphs with the same semantics and different syntaxes as test oracles to identify mis-compilation bugs in DL compilers.
The concept of constructing equivalent graphs/programs has shown potential in both DL framework testing and compiler testing~\cite{wang2022eagle,10.1145/3689757}.



\subsection{RQ3.2: Comparision and Analysis}
\label{sec:compiler_compare}

\begin{figure}
    \centering     
    \includegraphics[width=0.9\linewidth]{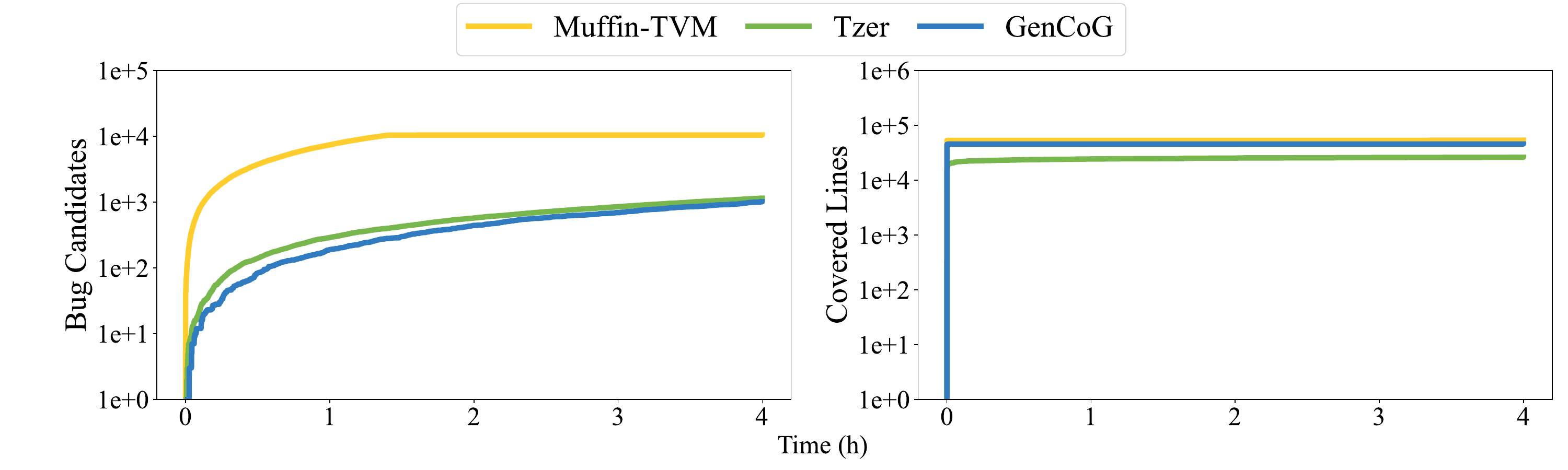}
    \caption{\updmn{Response to R3Q2: }{Comparison of Testing Methods on the TVM Compiler}}\label{fig:exptvm}
    \vspace{-10pt}
\end{figure}

\upd{Response to R1Q2 and R1Q6: }{{\bf Comparison Experiment and Results: }
We have compared three advanced open-sourced DL compiler testing methods to study their test effects, namely Tzer~\cite{liu2022coverage}, Muffin-TVM and GenCog~\cite{liu2023nnsmith}, where Muffin-TVM is a reimplementation of Muffin~\cite{gu2022muffin} by researchers, aiming to detect status bugs by converting DL models to relay IR with TVM’s Keras frontend.
We have followed their recommendation to repeatedly execute their source code for 4 hours in the same environment to separately detect bugs on TVM~\cite{liu2022coverage}.
Similar to the setting in~\autoref{sec:framedwork_compare}, in the experiments, we collected \ding{182} the code line coverage and \ding{183} the number of bug candidates generated by each method.
More details are in our repository~\cite{ourrepo}
\updmn{Response to R3Q2: }{The experimental results are shown in~\autoref{fig:exptvm}, where the X-axis represents time and the Y-axis represents line coverage and the number of bug candidates in scientific notation.}
We can observe that Tzer has the lowest line coverage, covering only 26.173 lines after four hours of execution, while the other two methods separately covered 45,610 lines and 53,309 lines within the first hour.
Particularly, Muffin-TVM covered more than 50,000 lines of TVM code and generated 10436 bug candidates in the experiment, significantly higher than the results of the other two methods.
Our analysis showed that such results could be related to its test case generation methods.
Muffin-TVM built diverse DL models using advanced framework testing methods and converted them to TVM for testing.
The model graphs carried rich information (\eg, operators) and conformed to the compiler's semantical constraints, thus effectively exploring DL compiler code lines and discovering bug candidates.
}

\updd{Response to R1Q2 and R1Q6: }{{\bf Analysis. }
Existing DL compiler testing methods mainly employ fuzzing techniques and generate test cases based on the compiler's semantical constraints.
These methods effectively explore various DL compiler behaviors while carrying useful information (\eg, operators and tensors) without violating the syntax of DL compilers, covering dozens of thousands of code lines in tests, as shown in~\autoref{fig:exptvm}.
In addition, both existing research~\cite{DBLP:journals/corr/abs-2407-16626} and our experiment results in~\autoref{fig:exptvm} demonstrate the potential of leveraging DL framework testing methods to cover a large amount of compiler code and trigger buggy cases in compiler testing.
However, existing methods have limitations in constructing test oracles, and can only detect limited bug types. 
They mainly rely on the state of program execution and different results between multiple DL compilers to identify crashes and inconsistency bugs, but can hardly identify optimization bugs and performance bugs in compilers.
For example, the Muffin-TVM can only detect crashes during TVM execution.
Although some researchers have proposed using metamorphic rules and equivalent graphs to construct test oracles, their methods are usually specific to limited operators and bug types and cannot cover diverse DL compiler operators and bugs, which limits the practical value of existing testing methods.
}


\section{DL Hardware Library Testing}
\label{sec:hardware}

In this section, we mainly focus on how existing work design testing methods for the DL hardware-related libraries (RQ2.3) and analyze their strengths and weaknesses (RQ3.3).
We have observed that existing research on DL hardware library testing mainly focuses on verifying and evaluating the correctness and efficiency, and rarely delves into bug detection with different testing techniques.
Therefore, we do not introduce existing work according to their testing techniques, but introduce the research on functional correctness and efficiency of DL hardware libraries respectively, and comparatively analyze the advantages and disadvantages of these studies in~\autoref{sec:hardwarelib_compare}.
\upd{Response to R3Q6: }{
We introduce representative testing methods in~\autoref{tab:hardware_compare} and explain how they overcome the above challenges in the following section.
Since we cannot find a specific description of the number and type of bugs, we do not show the relevant information in the table.}

\subsection{RQ2.3: DL Hardware Library Testing Methods}

\begin{table*}[]
    \caption{Representative DL Hardware Library Testing Methods}
    \label{tab:hardware_compare}
    \centering
    \scriptsize
    \tabcolsep=5pt
    \begin{tabular}{crr}
    \toprule
    Category & \multicolumn{1}{c}{Method Description} & \multicolumn{1}{c}{Test Libraries}\\
    \midrule
    \multirow{3}{*}{\begin{tabular}[c]{@{}c@{}}Testing on\\Functionality\end{tabular}} & \begin{tabular}[c]{@{}r@{}}Generated test patterns for DL accelerator to ensure\\ and reliability of its functional implementation~\cite{he2021efficient}\end{tabular} & NVDLA \\ \cmidrule{2-3}
    & \begin{tabular}[c]{@{}c@{}}Designed several MRs and conducted metamorphic testing \\ for the operators of DL accelerators~\cite{wang2020accuracy}\end{tabular} & HiAI/SNPE \\ \midrule
    \multirow{3}{*}{{\begin{tabular}[c]{@{}r@{}}Testing on\\Performance\end{tabular}}} & \begin{tabular}[c]{@{}r@{}}Conducted large-scale   experiments to evaluate \\ the performance of the convolution operators in cuDNN~\cite{jorda2019performance}\end{tabular} & cuDNN \\ \cmidrule{2-3}
    & \begin{tabular}[c]{@{}r@{}}Compared and evaluated the performance of the fixed \\ CNN architectures on three DL hardware libraries~\cite{nazir2023interpretable}\end{tabular} & \begin{tabular}[c]{@{}r@{}}cuBLAS, cuDNN,\\ and TensorRT\end{tabular} \\ \bottomrule
    \end{tabular}
\end{table*}

\subsubsection{Testing on DL Hardware Library Functional Correctness}

In terms of empirical research, Huang \et~\cite{huang2022demystifying} investigated and understood the dependency bug in the DL stack.
They investigated the symptoms and root causes of a total of 326 bugs in DL libraries which include DL applications, DL frameworks, DL accelerators, and DL hardware.
They found that violating the constraints among dependencies is the main root cause of bugs and then suggested that developers should receive systematic training to fully understand the DL stack and life cycle to reduce the occurrence of bugs.
Wang \et~\cite{wang2023compatibility} systematically studied the compatibility issues in the underlying DL libraries of DL systems that can affect the deployment of DL systems and lead to degradation of execution performance, including their symptoms and fixing patterns, providing insights into the testing and repairing of DL hardware libraries.
Their work promoted the following research on DL hardware-related libraries research and DL stack.

Researchers have also developed a series of methods to verify the correctness of the implementation of DL hardware libraries~\cite{he2021efficient,uezono2022achieving}.
He \et~\cite{he2021efficient} designed a set of functional testing methods for the compute units and control units of NVIDIA Deep Learning Accelerator (NVDLA) and achieved high test coverage and low test time overhead in tests.
However, their work relies on test coverage to verify the reliability of the library and did not construct test oracle to identify real-world bugs.
To construct effective test oracles in testing,
Wang \et~\cite{wang2020accuracy} conducted metamorphic testing and designed a series of MRs on the convolution and softmax operators of DL accelerators HiAI and Snapdragon Neural Processing Engine (SNPE) to verify the accuracy of accelerators and explore potential accuracy defects.
Their results showed that HiAI has better accuracy performance than SNPE on the float16 data type.
To explore the abnormal behavior of hardware libraries, researchers have leveraged fuzz testing to generate a variety of test cases.
CUDAsmith~\cite{jiang2020cudasmith} has been designed as a test case generation tool for the underlying NVCC and Clang library of the DL computing platform CUDA.
It implemented a generation-based kernel function generator to create valid test inputs that were adapted for the CUDA context and constructed test oracles based on random differential testing to identify bugs.
In their experiments, they mainly detected the build failures and timeout failures caused by wrong code.
All aforementioned test methods did not report identified real-world bugs in experiments.

\subsubsection{Testing on DL Hardware Library Efficiency}

Existing research mainly evaluates the efficiency of DL hardware libraries on specific operators or model architectures~\cite{jorda2019performance,nazir2023interpretable}.
\upd{Response to R1Q5: }{
Dongarra \et~\cite{dongarra2017design} conducted a large-scale evaluation on batched GEMM operation implemented in four DL hardware libraries, namely, MKL, OpenMP, CuBLAS, and MAGMA.
They found that the operation implemented in these libraries used simple memory layouts, leading to suboptimal performance.
Based on their findings, they proposed optimization that improves the performance of MKL and CuBLAS operations by up to 6 times.
Sun \et~\cite{sun2018evaluating} observed that existing work lacks attention to DL libraries on the ROC platform described by AMD.
To fill this gap and systematically evaluate the performance of DL libraries rocBLAS and MIOpen, Sun \et constructed four benchmarks, including K-Nearest Neighbors (KNN).
Their experiment results showed that MIOpen can perform well in both model training and inference processes.
Recently, {\"O}z \et~\cite{oz2024quantitative} conducted a detailed experimental analysis and evaluation of two BLAS libraries, cuBLAS and MAGMA on the basis of prior work.
The experiment results showed that CuBLAS and MAGMA did not exhibit significant differences in operations with large computational loads.
However, cuBLAS offered higher performance in terms of GFlops and achieved higher streaming multiprocessor utilization in most executions.
Beyond these DL libraries designed for computation and acceleration on hardware, Li \et~\cite{li2018tartan,li2019evaluating} conducted a systematic evaluation on four DL hardware libraries, PCIe, NVLink, NV-SLI, NVSwitch, and GPUDirect, which provide optimizations for GPUs interconnection.
They constructed empirical evaluations and observed that different GPU combinations have a significant impact on GPU communication performance.
For example, GPUDirect can exhibit the best performance on the supercomputer `Summit'.
Although the evaluation results of the aforementioned methods have provided valuable observations and insights on the efficiency of various DL hardware libraries, they still lacked testing case generation and further real-world bug detection.
}

\subsection{RQ3.3: Comparision and Analysis}
\label{sec:hardwarelib_compare}

\upd{Response to R1Q2: }{
Compared with the DL framework and compiler, existing testing methods on DL hardware libraries are relatively limited.
Researchers mainly built large-scale experiments to evaluate and verify the correctness and efficiency of specific DL hardware library functions.
In terms of efficiency, researchers have established large-scale experiments to evaluate the performance of DL hardware libraries in scenarios such as executing GEMM operators and inter-GPU communication.
However, existing performance evaluation lacks in-depth analysis and understanding of the root causes for poor performance.
Furthermore, there is no systematic study on the performance bug symptoms, root causes, and fix patterns of various DL hardware libraries.
In terms of correctness, although researchers have leveraged fuzzing and metamorphic testing to construct test cases and test oracles to verify the correctness of their functions, they can only evaluate and assess specific DL hardware libraries (\eg, CUDA accelerators), and have limited generalization ability, and also lack the ability to detect and identify real-world bugs.
}


\section{Future Gazing}
\label{sec:future}

\upd{Response to R1Q2, R1Q3 and R2Q3: }{
Although the testing research on the DL library has made considerable progress, this field is still in its nascent stage.
There are still many challenges to be addressed.
In this section, we first conclude the findings from the prior literature study, aiming to explore practical implications and provide meaningful conclusions.
Subsequently, based on these findings, we list the challenges faced by existing methods and potential research opportunities, aiming to provide insights and a clear set of guidelines for future research in DL library testing.
}

\subsection{RQ4.1: Findings}




\noindent
\(\bullet\)
\upd{Response to R1Q3 and R2Q3: }{{\bf Finding 1: }{\it Conducting empirical studies can promote the design of effective DL library testing methods.}
Empirical studies can provide valuable findings of DL library bugs, root causes, and fix patterns and help the following research construct effective testing methods, which have been shown in related literature~\cite{chen2022toward,shen2021comprehensive}.
In addition, with the continuous evolution of DL libraries, it is meaningful to study and summarize the distribution and fix patterns of bugs in their latest releases.
For example, early studies on DL frameworks focused on bugs in programs using a specific framework and lacked analysis of the fix patterns~\cite{zhang2018empirical}.
Recent researchers have analyzed thousands of bugs in four frameworks, as well as their root causes and fix patterns, providing more comprehensive and detailed findings on DL framework testing than earlier studies~\cite{chen2022toward}.
However, existing research lacks of study on bugs of various DL hardware libraries (\eg, MKL).
Systematically studying bugs in various DL hardware libraries and pointing out their root causes is of practical implication for developing new DL library testing methods.
}

\noindent
\(\bullet\)
\updd{Response to R1Q3 and R2Q3: }{{\bf Finding 2: }{\it DL library testing research mainly focuses on DL frameworks at the high abstraction level and pays less attention to the underlying DL compiler and hardware libraries.}
DL compilers and hardware libraries optimize the abstract DL models and execute corresponding operators on specific hardware.
The bugs in these libraries can directly affect the calculation results of DL models.
Even worse, they can hardly be detected in higher-level tests (\eg, tests on DL systems and DL frameworks).
However, researchers have paid insufficient attention to testing existing DL compilers and hardware libraries.
Among the papers we collected, only 12.90\% and 13.98\% are related to DL compilers and DL hardware library testing, respectively.
Especially for DL hardware libraries, the related literature lacked systematic and comprehensive testing methods to identify real-world bugs~\cite{nazir2023interpretable,oz2024quantitative}, which limits the practical value of these methods.
}

\noindent
\(\bullet\)
\updd{Response to R1Q3 and R2Q3: }{{\bf Finding 3: }{\it Although researchers have proposed various DL framework testing methods, it is difficult to compare their test effects in real-world scenarios.}
Based on the collected literature, we have observed that, in the recent five years, dozens of DL framework testing methods and tools have been proposed, which conducted testing for different bugs such as crashes, inconsistent output, NaN, performance problems, and documentation bugs~\cite{guo2020audee,wei2022free,xie2022docter}.
However, it is difficult to compare the effectiveness of these methods in real-world scenarios.
On the one hand, different testing methods discover different types of bugs (\eg, performance bugs and crashes) on different library releases, making it difficult to quantitatively compare these testing methods.
On the other hand, existing methods often output thousands of bug candidates (\autoref{fig:expframework}).
Verifying and determining real bugs requires a significant amount of manual effort.
Although researchers have compared several test tools based on metrics like line coverage~\cite{chen2022toward}, and we have also conducted comparative experiments for five state-of-the-art methods in~\autoref{sec:framedwork_compare}, there is still a lack of comprehensive benchmarks to compare and evaluate the effectiveness of existing DL framework testing methods.
Such benchmarks are essential not only to create standardized baselines for future research but also to encourage broader deployment and practical application of various testing methods in real-world scenarios.
}

\noindent
\(\bullet\)
\updd{Response to R1Q3 and R2Q3: }{{\bf Finding 4: }{\it Existing testing and evaluation on DL libraries mainly focuses on functionality correctness (\eg, crashes) and pays less attention to performance bugs.}
Performance bugs can introduce significant runtime and resource overhead to DL libraries and systems built upon them, resulting in energy waste and environmental concerns in real-world deployment~\cite{nistor2015caramel,lacoste2019quantifying}.
However, in our literature study, only 38.71\% of collected papers paid attention to performance bugs, and most of them only conducted empirical studies on performance bugs.
Even though several methods claimed to be able to detect performance bugs, they usually involved identifying one or two cases of memory leaks or performance degradation, lacking systematic testing methods~\cite{wei2022free,zhang2024citadel}.
There is an urgent need to discover performance bugs and ensure the efficiency of DL libraries and even DL systems in real-world scenarios.
}


\noindent
\(\bullet\)
\updd{Response to R1Q3 and R2Q3: }{{\bf Finding 5: }{\it Combining and leveraging the advantages of multiple testing techniques has the potential to bring better testing results to DL library testing.}
Using a single testing technique can only cover specific bug types (\eg, crashes~\cite{shi2023acetest}), while combining multiple techniques has the potential to detect diverse bug types and cover a variety of DL library APIs, which has been demonstrated in related literature~\cite{wei2022free,deng2023large}.
In contrast, DL compiler and hardware library research typically relied on simple testing and evaluation methods, focusing on specific bugs like mis-compilation~\cite{xiao2022metamorphic}.
We suggest that future developers can combine various techniques (\eg, metamorphic testing and ML techniques) to generate valid test cases and construct accurate test oracles, thereby effectively identifying various bugs in real-world scenarios.
}

\noindent
\(\bullet\)
\upd{Response to R1Q3, R2Q3, and R3Q2: }{{\bf Finding 6: }{\it The DL library testing method mainly detects bugs and reports them to developers, rarely discovering security vulnerabilities.}
In the literature collection, we have observed that most existing works focus on DL library bugs that degrade the functionality and performance of DL libraries.
Only a small number of research tests and discovers security vulnerabilities and weaknesses in DL libraries~\cite{christou2022ivysyn,lai2024security,zhu2025model}.
The vulnerabilities in DL libraries can be exploited by attackers, leading to memory exhaustion, kernel crashes, etc. that compromise the security of DL-based software and systems.
With the widespread application of DL technology, we suggest that industry and academia pay more attention to the security vulnerabilities and weaknesses in DL libraries.
}

\noindent
\(\bullet\)
\upd{Response to R1Q3 and R2Q3: }{{\bf Finding 7: }{\it The continuous evolution of DL techniques poses challenges for the maintenance and testing of DL libraries.}
Existing work on DL library testing often focuses on identifying bugs in specific versions of the library.
However, libraries continue to have new versions released, which may fix some inherent bugs while potentially introducing new ones.
Regression bugs are also huge pitfalls in DL libraries\footnote{https://github.com/pytorch/pytorch/issues/95432}, where previously functional functions in older versions become problematic in the new version.
In addition, DL libraries that emerged in the era of LLMs (\eg, HuggingFace Transformers~\cite{wolf-etal-2020-transformers}) may also introduce risks to DL workflow and systems.}
\upd{Response to R3Q12: }{
However, existing research~\cite{jia2021unit} revealed that the unit test cases in the DL libraries are not sufficient to discover potential bugs.
To effectively detect bugs in the ever-changing DL library,
existing work~\cite{xie2024cedar} has proposed continuous testing methods to test different releases of libraries, but they focus on testing different releases directly with existing techniques and lack attention to the characterization of the evolution of the frameworks.
Building a lightweight bug benchmark for various DL libraries is of practical significance in effectively identifying bugs during software evolution, which can reduce the cost of maintaining and developing DL libraries.}

\subsection{RQ4.2: Challenges and Opportunities}
Based on the above findings, we summarize the main challenges and the research opportunities as follows.

\noindent
\(\bullet\)
\upd{Response to R1Q3: }{{\bf Challenge 1: Collecting bug reports to study various DL Hardware libraries.}
The major challenge in conducting empirical research for DL hardware library bugs is collecting various bug cases.
Different from open-sourced DL frameworks and compilers, DL hardware libraries are usually designed for specific hardware and do not provide available code or bug (issue) lists.
In addition, there are various DL hardware libraries covering different scenarios such as computing and hardware communication.
Researchers could first collect buggy cases or projects of a specific hardware library from open-source communities such as Stack Overflow, proposing a taxonomy for these bugs.
Then they could extend the research to different libraries, providing valuable findings for DL hardware library testing.
}

\noindent
\(\bullet\)
\upd{Response to R3Q9, and R3Q11: }{{\bf Challenge 2: Constructing test cases and oracles for DL compiler and hardware library.}
There are two major challenges in testing and identifying bugs in DL compilers and hardware libraries
The first challenge lies in building valid test cases.
The inputs of DL compilers and hardware libraries often consist of specific programming languages and operators, which makes it difficult to find enough usable reference code and test cases in the open-source community and extract input constraints.
Building test oracles for these libraries is another challenge.
Especially for DL hardware libraries designed for specific hardware, it is difficult to build test oracles for them using different devices or implementations.
Constructing equivalent graphs or generating test cases based on LLM techniques which have shown strong capabilities in traditional SE testing tasks (\eg, guiding differential and fuzz testing~\cite{etemadi2024mokav,hossain2024togll,deng2023large}), could be a potential solution to overcoming these challenges.
In addition, our experiment results in~\autoref{fig:exptvm} and existing research~\cite{DBLP:journals/corr/abs-2407-16626} show that using advanced DL framework testing methods to effectively generate diverse test cases and convert them into underlying libraries is also a promising research direction.
}

\noindent
\(\bullet\)
\upd{Response to R3Q11: }{{\bf Challenge 3: Conducting a comprehensive evaluation of DL framework testing methods.}
Firstly, a major challenge in constructing benchmarks for existing testing methods is to collect a comprehensive dataset including various bugs on different DL frameworks.
DL frameworks evolve rapidly, and various bugs typically exist in the different releases and environments.
Researchers could refer to existing software testing datasets and benchmarks~\cite{just2014defects4j,ben2019modular,li2020benanza,liang2022gdefects4dl} to build a dataset that covers a large number of APIs and various types of bugs and use virtual environment management tools like Docker, thus supporting bug datasets for multiple DL frameworks and releases.
Another challenge lies in automated processing output results of existing testing methods (\ie, bug candidates) to obtain the real bugs they detected.
Existing research often reports thousands of bug candidates (as discussed in~\autoref{sec:framedwork_compare}), and evaluating and verifying the detection results of detection methods requires a significant amount of manual effort.
Automatically merging bug candidates based on ML techniques is a potential research direction.
}

\noindent
\(\bullet\)
\upd{Response to R3Q11: }{{\bf Challenge 4: Identifying performance bugs on DL libraries.}
The major challenge in identifying performance bugs is designing effective test oracles to estimate the expected performance of a given DL program on the hardware.
Existing research primarily leverages metamorphic relations or prior bug reports to construct test oracles and detect performance bugs.
For example, researchers compare the executions of the same function with different data types and observe whether the memory or time overhead has increased or decreased as expected, which could provide a qualitative test oracle~\cite{wei2022free}.
However, such a method can not precisely predict the performance of the given API or operator quantitatively, and can only identify the performance bugs related to data types.
}

\noindent
\(\bullet\)
{\bf Challenge 5: Designing effective DL library testing methods with multiple testing techniques.}
How to combine the advantages of different testing techniques to design new testing methods is a challenge.
As we analyzed in~\autoref{sec:framedwork_compare}, different testing techniques have their strengths and limitations.
For example, differential testing can provide general test oracles for diverse APIs, but it may lead to false positives in tests.
Fuzzy testing can efficiently generate test samples.
Researchers can refer to previous methods~\cite{wei2022free,deng2023large}, using fuzzing and ML techniques to efficiently generate test cases to cover behaviors of different DL libraries and designing some general MRs and equivalent relationships to construct accurate test oracles for various bugs.

\noindent
\(\bullet\)
\upd{Response to R3Q2: }{{\bf Challenge 6: Discovering security vulnerabilities of DL libraries.}
Discovering vulnerabilities in DL workflow and building reasonable threat models is the major challenge.
On the one hand, researchers can generate test cases based on the security properties summarized in existing machine learning security research~\cite{zhang2020empirical,zhang2022testing} to evaluate and identify problems in the DL library.
On the other hand, researchers can also explore potential weaknesses and vulnerabilities based on common attack scenarios in AI security and software security, combined with the characteristics of different DL library components, such as using DL library APIs to perform unauthorized operations, illegally modifying system files of host, and implanting backdoors into DL software and systems.
}

\noindent
\(\bullet\)
\upd{Response to R3Q11 and R3Q12: }{{\bf Challenge 7: Designing benchmarks to test DL libraries during evolution.}
The major challenge in designing bug benchmarks for DL libraries is how to efficiently design test cases for evolving libraries, thereby covering the diverse functionalities of the libraries.
As new optimization algorithms and model architectures are continuously proposed and implemented~\cite{zhangrevisiting,liumobilellm}, updating and modifying test cases in the benchmark requires a lot of manual effort.
A promising approach is to leverage code similarity metrics and Software Component Analysis (SCA) to identify existing components in the DL library that are functionally similar to newly released components and migrate the test cases of the former to the latter, thereby reducing the effort of maintaining the benchmark.
}

\section{Conclusion}
\label{sec:conclusion}


The rapid development and widespread deployment of DL-driven systems have attracted researchers in academia and industry to investigate and study the DL underlying library that supports DL systems.
Existing research has achieved fruitful results in testing and validating DL library bugs.
However, with the development and iteration of DL techniques and software, there is still room for improvement in these testing methods.
To comprehensively summarize the testing research of DL underlying libraries, understand their effectiveness and limitations, and discuss challenges and directions for future research, this paper first describes the definitions of DL library bugs and testing.
Then, based on four proposed review questions, it reviews the existing testing research on three components of DL libraries (\ie, DL frameworks, DL compilers, and DL hardware libraries).
Finally, this paper summarizes the findings of the literature survey and discusses the challenges of DL library testing,  aiming to promote further development and real-world application of DL library testing research.


\begin{acks} 
The authors would like to thank the anonymous reviewers for their insightful comments and valuable suggestions.
This work is supported partially by the National Key Research and Development Program of China (2023YFB3107400), the National Natural Science Foundation of China (62006181, 62132011, 62161160337, 62206217, U20A20177, U21B2018), and the Shaanxi Province Key Industry Innovation Program (2021ZDLGY01-02 and 2023-ZDLGY-38).
Thanks to the New Cornerstone Science Foundation and the Xplorer Prize.
\end{acks}


{ \scriptsize
\bibliographystyle{ACM-Reference-Format}
\bibliography{sample-base}
}




\end{document}